\documentclass{aa} 
\usepackage{graphicx,natbib,psfig,amssym} 
\newcommand{\ltsima} {$\; \buildrel < \over \sim \;$} 
\newcommand{\gtsima} {$\; \buildrel > \over \sim \;$} 
\newcommand{\lta} {\lower.5ex\hbox{\ltsima}} 
\newcommand{\gta} {\lower.5ex\hbox{\gtsima}}

\newcommand{\ergscm}{\>{\rm erg}\,{\rm s}^{-1}\,{\rm cm}^{-2}}
\newcommand{\ergscmA}{\>{\rm erg}\,{\rm s}^{-1}\,{\rm cm}^{-2}\,{\rm \AA}^{-1}}
\newcommand{\kms}{$\rm{\,km \,s}^{-1}$}
\begin{document} 
\title{Recent star formation in nearby 3CR radio-galaxies\\ 
  from UV HST observations
\thanks{Based  on observations obtained at
the  Space  Telescope Science  Institute,  which  is  operated by  the
Association of  Universities for Research  in Astronomy, Incorporated,
under NASA contract NAS 5-26555.}}
  
\titlerunning{Recent star formation in nearby 3CR radio-galaxies}   

\authorrunning{R.D. Baldi \& A. Capetti}
  
\author{Ranieri D. Baldi
\inst{1}
\and  Alessandro Capetti \inst{2}} 
\offprints{R.D. Baldi}  
\institute{
Universit\'a di Torino, via P. Giuria 1, 10125 Torino, Italy\\
\email{baldi@oato.inaf.it}
\and 
INAF - Osservatorio Astronomico di Torino, Strada
  Osservatorio 20, I-10025 Pino Torinese, Italy\\
\email{capetti@oato.inaf.it}}

\date{}  
   
\abstract{We analyzed HST images of 31 nearby (z $\lesssim$ 0.1) 3CR
  radio-galaxies.  We compared their UV and optical images to detect evidence
  of recent star formation. Six objects were excluded because they are highly
  nucleated or had very low UV count rates.  After subtracting the emission
  from their nuclei and/or jets, 12 of the remaining 25 objects, presenting an
  UV/optical colors NUV - r $<$ 5.4, are potential star-forming candidates.
  Considering the contamination from other AGN-related processes (UV emission
  lines, nebular continuum, and scattered nuclear light), there are 6
  remaining star-forming ``blue'' galaxies.
 
We then divide the radio galaxies, on the basis of the radio morphology, radio
power, and diagnostic optical line ratios, into low and high excitation
galaxies, LEG and HEG. While there is no correlation between the FR type (or
radio power) and color, the FR type is clearly related to the spectroscopic
type. In fact, all HEG (with one possible exception) show morphological
evidence of recent star formation in UV compact knots, extended over 5-20 kpc.
Conversely, there is only 1 ``blue'' LEG out of 19, including in this class
also FR~I galaxies.

The picture that emerges, considering color, UV, optical, and dust morphology,
is that only in HEG recent star formation is associated with these relatively
powerful AGN, which are most likely triggered by a recent, major, wet merger.
Conversely, in LEG galaxies the fraction of actively star-forming objects is
not enhanced with respect to quiescent galaxies. The AGN activity in these
sources can be probably self-sustained by their hot interstellar medium.

\keywords{Galaxies: active -- Galaxies: elliptical and lenticular, cD -- Galaxies:
  photometry -- Galaxies: evolution -- Galaxies: interactions -- Galaxies:
  starburst -- Ultraviolet: galaxies}} 

\maketitle
  
\section{Introduction}
\label{introduction}

Both observational and theoretical studies have supported the idea of a
co-evolution between supermassive black holes (SMBH) and their host
galaxies. The observational evidence is derived from several pieces of
evidence: the dynamical signature of the widespread presence of SMBH in
galaxies (e.g \citealt{kormendy95}; \citealt{richstone98};
\citealt{kormendy01}), the relation between SMBH mass and the spheroid mass
(\citealt{magorrian98}; \citealt{mclure02}; \citealt{marconi03};
\citealt{haring04}), stellar velocity dispersion (\citealt{ferrarese00};
\citealt{gebhardt00}; \citealt{tremaine02}) and concentration index
(\citealt{graham01}; \citealt{graham07}). These relations were interpreted by
e.g. \citet{hopkins07c} as various projections of a fundamental plane, which
included the SMBH mass, analogous to a similar relation for elliptical
galaxies (\citealt{dressler87}; \citealt{djorgovski87}); this appears to be a
sign of a common evolutionary process involving SMBH and galaxies.

The growth of a SMBH can occur via either gas accretion or coalescence with
another SMBH during a merger event. Similarly, the growth of a galaxy can be
associated with the capture of the stellar populations of another galaxy, due
to either a merger or to star formation, which might be self-sustained or
triggered by a merger.  In this framework, nuclear activity (i.e. the
manifestation of gas accretion onto a SMBH) and star formation are expected to
be related, and for both processes mergers are likely to play a crucial role.
Indeed, the hierarchical galaxy evolution scenarios support the idea that gas
flows associated with galaxy mergers trigger both starburst and AGN activity
(\citealt{kauffmann00}; \citealt{dimatteo05}). A connection between mergers,
galactic starburst, and AGN activity has been well-established and modeled
(e.g. \citealt{barnes91,barnes96}; \citealt{mihos94,mihos96};
\citealt{springel05}; Kapferer et al. 2005). \citet{hopkins06a} presented a
simulation in which starburst, AGN activity, and SMBH growth were connected by
an evolutionary sequence, due to mergers between gas-rich
galaxies. Observationally, studies of ultraluminous infrared galaxies
(ULIRGs), distant submillimeter galaxies (SMGs), and quasars demonstrate that
they are the remnants of major mergers, in which massive starbursts occur in
combination with a dust enshrouded AGN (\citealt{barnes96};
\citealt{schweizer98}; \citealt{jogee06}; \citealt{sanders88a,sanders88c};
\citealt{sanders96}; \citealt{dasyra07}).  Studies of QSO host galaxies also
revealed the widespread presence of a young stellar population
(e.g. \citealt{brotherton99}; \citealt{kauffmann03}; \citealt{sanchez04};
\citealt{zakamska06}). In Seyfert galaxies, evidence of nuclear, dusty,
compact starbursts was found (\citealt{heckman97}; \citealt{gonzalez98}).
Apparently, AGN activity and star formation are connected also from a
quantitative point of view, since the most luminous quasars have the youngest
host stellar populations (\citealt{jahnke04a}; \citealt{vandenb06}) and the
most significant post-merger tidal features and disturbances
(e.g. \citealt{canalizo01}; \citealt{hutchings03}; \citealt{letawe07}).

Most of studies focused on the general AGN population when radio-quiet AGN are
not studied in isolation. However, there have been plentiful studies of
analysis of radio-loud AGN, which are the focus of this paper.  Observations
of radio galaxies in the local Universe (\citealt{heckman86};
\citealt{smith89}; \citealt{colina95}) detected morphological features such as
double nuclei, arcs, tails, bridges, shells ripples, and tidal plumes. In
a substantial subset of radio galaxies, this suggests that the AGN activity is
triggered by the accretion of gas during major mergers and/or tidal
interactions. Furthermore, gas kinematics studies (\citealt{tadhunter89} and
\citealt{baum92}) supported the interpretation that mergers can represent the
triggering process of SMBH/galaxies co-evolution. Studies of the spectral
energy distribution (SEDs) of radio galaxies discovered that young stellar
populations, indicative of a recent starburst, provide a significant
contribution to the optical/UV continua, up to 25-40 \%, of powerful galaxies
at low and intermediate redshift (\citealt{lilly84}; \citealt{smith89};
\citealt{tadhunter96}; \citealt{melnick97}; \citealt{aretxaga01};
\citealt{odea01}; \citealt{tadhunter02}; \citealt{wills02}; \citealt{allen02};
\citealt{wills04}; \citealt{raimann05}; \citealt{tadhunter05};
\citealt{holt07}).  Apparently, AGN activity initiates at a later stage of the
merger event than star formation \citep{emonts06}, which agrees with
predications of numerical simulations \citep{hopkins07}, although the precise
length of the delay is not well constrained.

However, there are still several open questions, such as: (i) what is the
fraction of nearby radio galaxies showing evidence for young stars?; (ii) is
there a relationship between AGN properties and galaxy star-formation
characteristics?; (iii) what is the role of mergers in triggering both AGN
activity and star formation in radio-galaxies?

We consider these issues by adopting an alternative approach based on the
analysis of UV images with the aim of detecting a young stellar population.
More specifically, we analyzed HST/STIS observations of a sample of nearby
radio galaxies from the 3CR catalogue. The flux level in the UV band ($\sim$
2500 \AA), relative to the optical emission, is sensitive to very low levels
of star formation, as demonstrated by GALEX data (the GALEX/NUV band is
similar to the filter used to acquire the STIS images used in this study),
by detecting fraction of as low as 1-3 \% of young stars formed within
the last billion years (\citealt{silk77} and \citealt{somerville00}). The HST
UV images were already used for this purpose by \citet{odea01}, who studied
the radio-galaxy 3C~236. The sample considered is sufficiently large (31
objects) for statistical conclusions and is representative of the overall
population of nearby radio galaxies.

Furthermore, \citet{schawinski06} recently analyzed GALEX and SDSS images of a
significant number of non-active early-type galaxies. This study provides a
well defined control sample that we can use as a benchmark for the star
formation properties of quiescent galaxies to be compared with the
measurements we will derive for radio-galaxies. We note that
\citet{schawinski06} focused only on quiescent galaxies to avoid contamination
by UV light from an active nucleus.  For the sample that we consider here, the
light produced by AGN activity (UV nuclei, and jets) can however be masked by
taking advantage of the high resolution of the HST images.

The outline of this paper is the following. In Sect. \ref{observations and
data reduction}, we describe the properties of the sample and HST
observations, on which this study is based, as well as the reduction method
used. In Sect. \ref{results}, we present our results by classifying the
galaxies in terms of their NUV-r color profiles and UV morphology.  In
Sect. \ref{discussion}, we discuss the link between AGN and star formation
properties, while in Sect. \ref{summary} we summarize our findings.  The
cosmological parameters used in this paper are H$_{0}$ = 75 \kms\ Mpc$^{-1}$
and q$_{0}$ = 0.5.
 
\section{Observations and data reduction}
\label{observations and data reduction}

We selected all 3CR radio-galaxies for which STIS near-UV observations were
available in the public archive.  The sample consists of 31 low-z
radio-galaxies with z $<$ 0.1, apart form 3C~346 at z = 0.16.
For 25 objects, the UV data were published by \citet{allen02}. To this
sample we added a subset of 6 objects observed by various HST projects.
Our final sample represents $\sim$60 \% of the entire 3CR sample with z
$<$ 0.1 (49 galaxies) that contains an almost equal fraction (and coverage) of FR~I and
FR~II galaxies. Since the objects discussed in \citet{allen02} sample were
selected at random (since the data were obtained as part of a HST snapshot
proposal) and only 6 objects were added our sample is representative of
the entire low-z 3CR sample and has no evident selection biases.

We analyzed UV and optical HST images of our sample.  All UV observations
employed the STIS near-ultraviolet (NUV) Cs$_{2}$Te Multianode Microchannel
Array (MAMA) detector, which has a field of view of
25$^{\prime\prime}\times$25$^{\prime\prime}$ and a pixel size of $\sim
0\farcs$024. The filters used during observations were the following: most
objects were observed with the F25SRF2 filter (centered on 2320 \AA\ with a
FWHM of 1010 \AA); for three additional galaxies, the F25CN182 filter was used
(with a pivot wavelength of 1983 \AA\ and a FWHM of 630 \AA), while the
remaining 5 observations were obtained with the F25QTZ filter (centered on
2364.8 \AA\ with a FWHM of 995.1 \AA), which provides more effective rejection
of the OI$\lambda$1302 \AA\ geocoronal emission than F25SRF2.  All of these
filter band cutoffs remove geocoronal Ly$\alpha$ emission. The observations
log is given in Table \ref{tab1}.

The optical observations were acquired using the WFPC2 camera.  Most objects
were located in the PC field with a field of view of $\sim 36\farcs4 \times
36\farcs4$ and pixel size of $\sim0\farcs$04553. Some objects were located in
the WF fields, which individually have a field of view of
1.3$^{\prime}\times$1.3$^{\prime}$ and pixel size of $\sim 0\farcs1$.  For all
galaxies, the used filter was F702W with a pivot wavelength of 6919 \AA\ and a
bandwidth of 1385 \AA\ \footnote{Only one object (3C~192) lacked the WFPC2
images and then its SDSS optical image has been employed}.  The WFPC2 exposure
times were 280 s or 560 s. The optical observations were published by
\citet{dekoff96} and \citet{martel99}.

The UV and optical images were processed by the standard HST pipeline,
developed at the STScI, in the package STSDAS in IRAF\footnote{IRAF is
distributed by the National Optical Astronomical Observatories, which are
operated by the Association of Universities for Research in Astronomy, Inc.,
under cooperative agreement with the National Science Foundation.} which
corrects for flat-fielding and, for optical images, also bias subtraction. For
optical data, the CCD images were combined and the cosmic-ray events were
cleaned in a single step using the IRAF/STSDAS task crrej, when multiple
exposures of the same target were available; otherwise cosmic rays were
removed using the task cosmicrays of the IRAF package.
   
The images were flux calibrated by using the keyword PHOTFLAM in the
calibrated science header file, and divided by the value of the keyword
EXPTIME (exposure time). The expected calibration error was given by the error
in the value of PHOTFLAM: for optical band images, this was 2 \% and, for the
UV band, this was 5 \%.

\subsection{UV / optical colors}

To estimate the UV / optical colors, we measured the total UV and optical flux
for each object in the sample.  The optical images were registered onto the UV
images; the registration was performed after re-binning the UV images to a
pixel size of 0$\farcs$45 as to enable detection of regions of low brightness
UV emission. We registered the images by aligning, when present in the field
of view, optical and UV nuclei and other objects (companion galaxies and/or
stars). Otherwise, in the case of diffuse sources (e.g. 3C~40), we
cross-correlated the UV with the R images to estimate the relative shift.

Different masks, defined individually for each galaxy, were applied to both
images, which identified areas containing nearby galaxies or stars. More
importantly, we flagged regions contaminated by emission from extended jets.
The presence of a optical and/or UV jet is noted in the Table \ref{tab1}.
Similarly, we masked the nuclear regions (with a typical size of
0$\farcs$12-0$\farcs$25) in all objects where the studies of
\citet{chiaberge99} and \citet{chiaberge02} identified the presence of
unresolved nuclei in the optical and/or UV images (as marked in Table
\ref{tab1}).

The background level was estimated from a circular annulus of 12$\farcs$5 in
radius and 3$\farcs$6 in width for both the optical and UV images. This was
the largest region in the UV images that had not been contaminated by enhanced
dark current at the edges of the STIS detector. For objects located close to
the edges of the frame, we used instead circular sectors for the background
estimate.  The UV background rate measurements were compared with those
obtained from the same images by \citet{allen02}: our values were in good
agreement with their estimates, although they are lower on average by 20\%
(and as much as 40 \% for one object). This is most likely because the region
that we used to measure the background was located at larger radii from the
center of the galaxy.

We measured UV and optical fluxes within a set of concentric apertures,
centered on each object, with different radii 2$^{\prime\prime}$,
4$^{\prime\prime}$, 6$^{\prime\prime}$, 8$^{\prime\prime}$, and
10$^{\prime\prime}$. From the total counts within each region, we subtracted
the estimated background and applied the appropriate photometric conversion
to obtain a flux in units of erg s$^{-1}$ cm$^{-2}$ \AA$^{-1}$. These
fluxes were then transformed onto the AB magnitude system and used to estimate
the NUV-R color.  With this approach, we measured the integrated NUV-R color
integrated over different apertures out to the distance at which the error in the
color was smaller than 0.2 mag\footnote{The error in the NUV-R color increases
with increasing aperture due to the uncertainty in the UV background.}.

At this stage, we discarded 6 galaxies of the sample.  For two objects (namely
3C~353 and 3C~452), the integrated UV count rate was insufficient to measure
accurately (with an error smaller than 0.5 mag) the galaxy's color.  Four
targets (namely 3C~227, 3C~371, 3C~382, and 3C~390.3) were highly
nucleated. The color of these galaxies were strongly dependent on the nuclear
subtraction and so it was difficult to separate any genuine diffuse emission
from the halo of the bright nuclear point source that might dominate even at
significant distances from the galaxy center.

\begin{figure*}
\centerline{
\psfig{figure=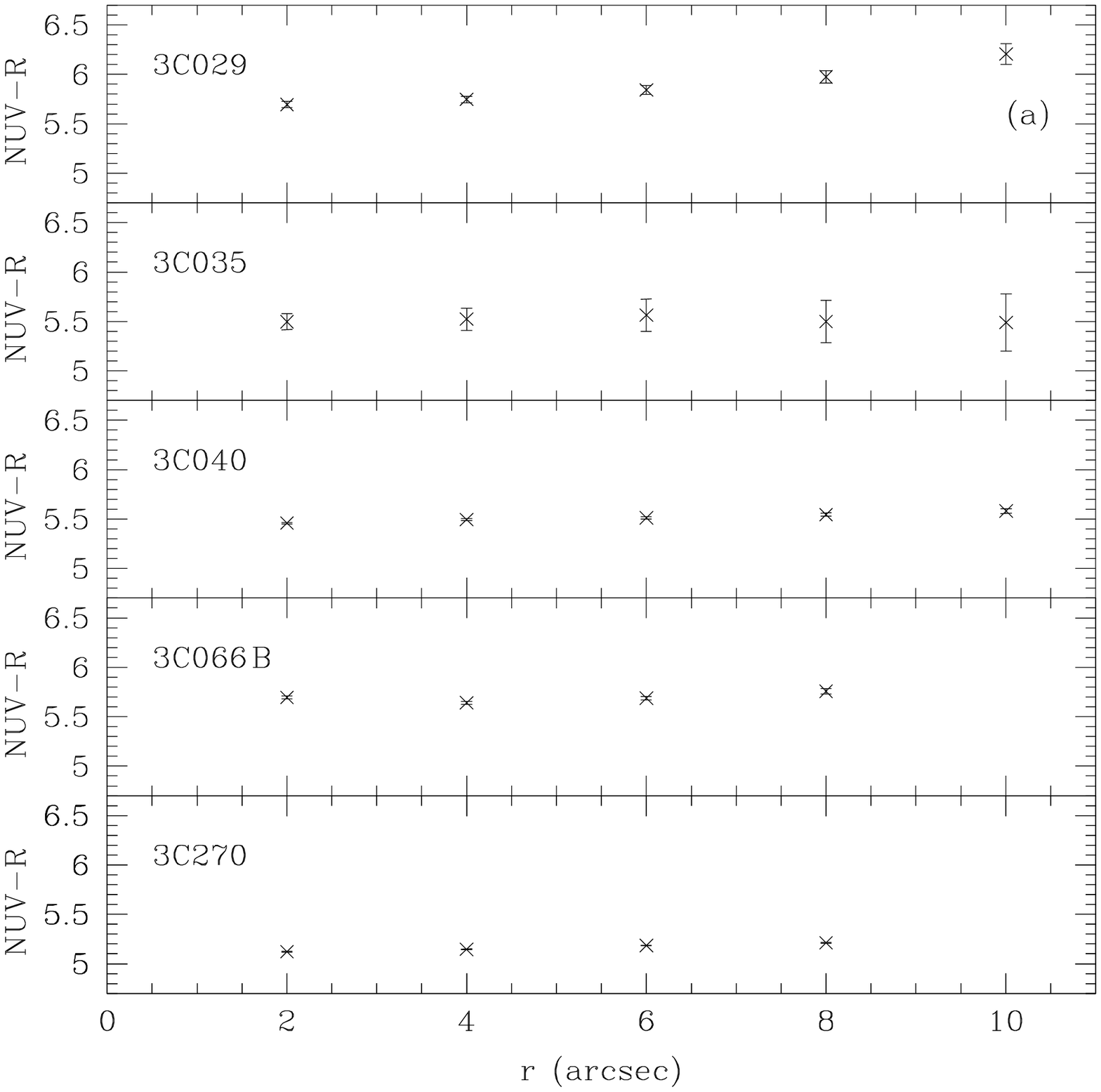,width=0.33\linewidth}
\psfig{figure=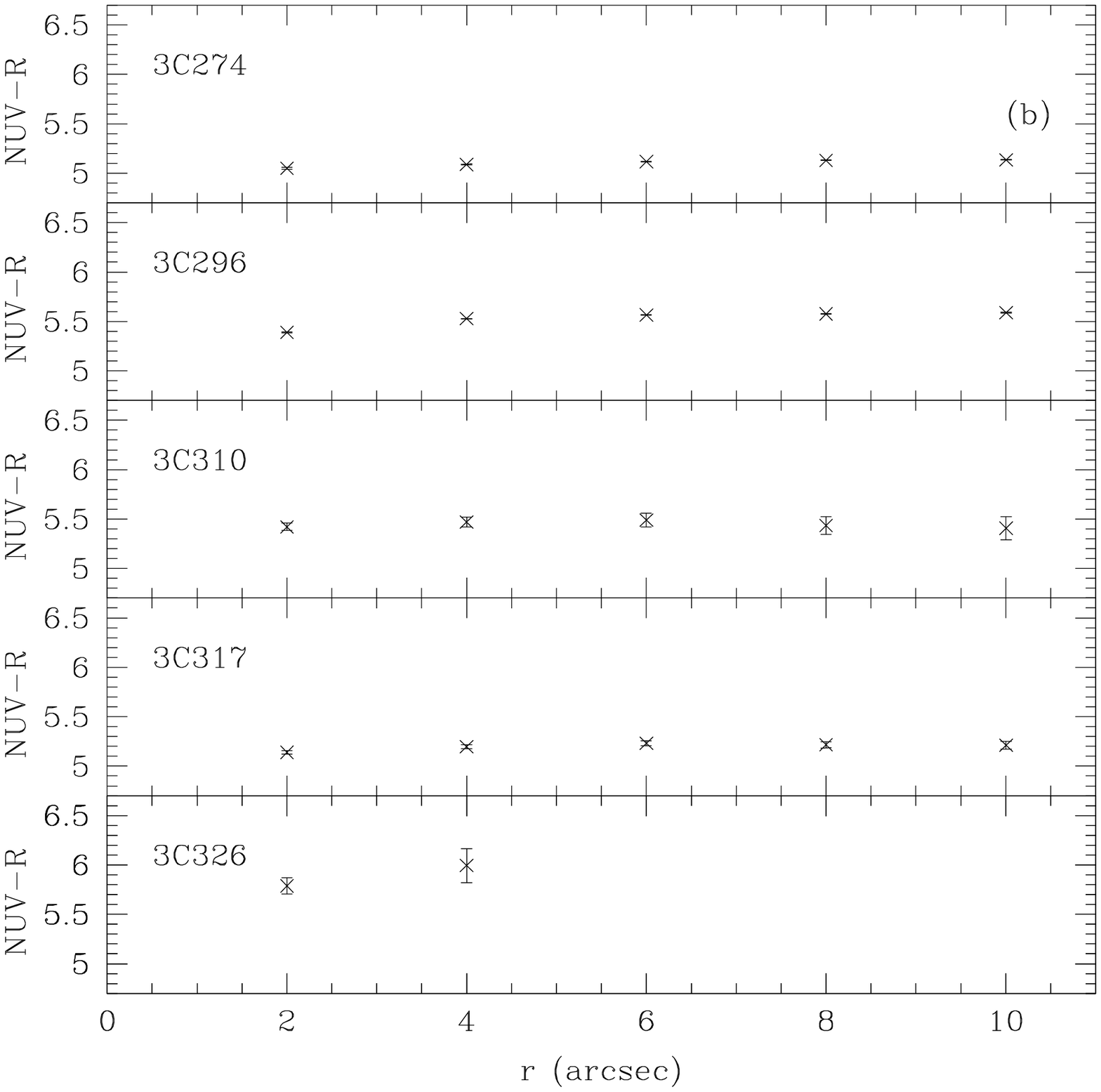,width=0.33\linewidth}
\psfig{figure=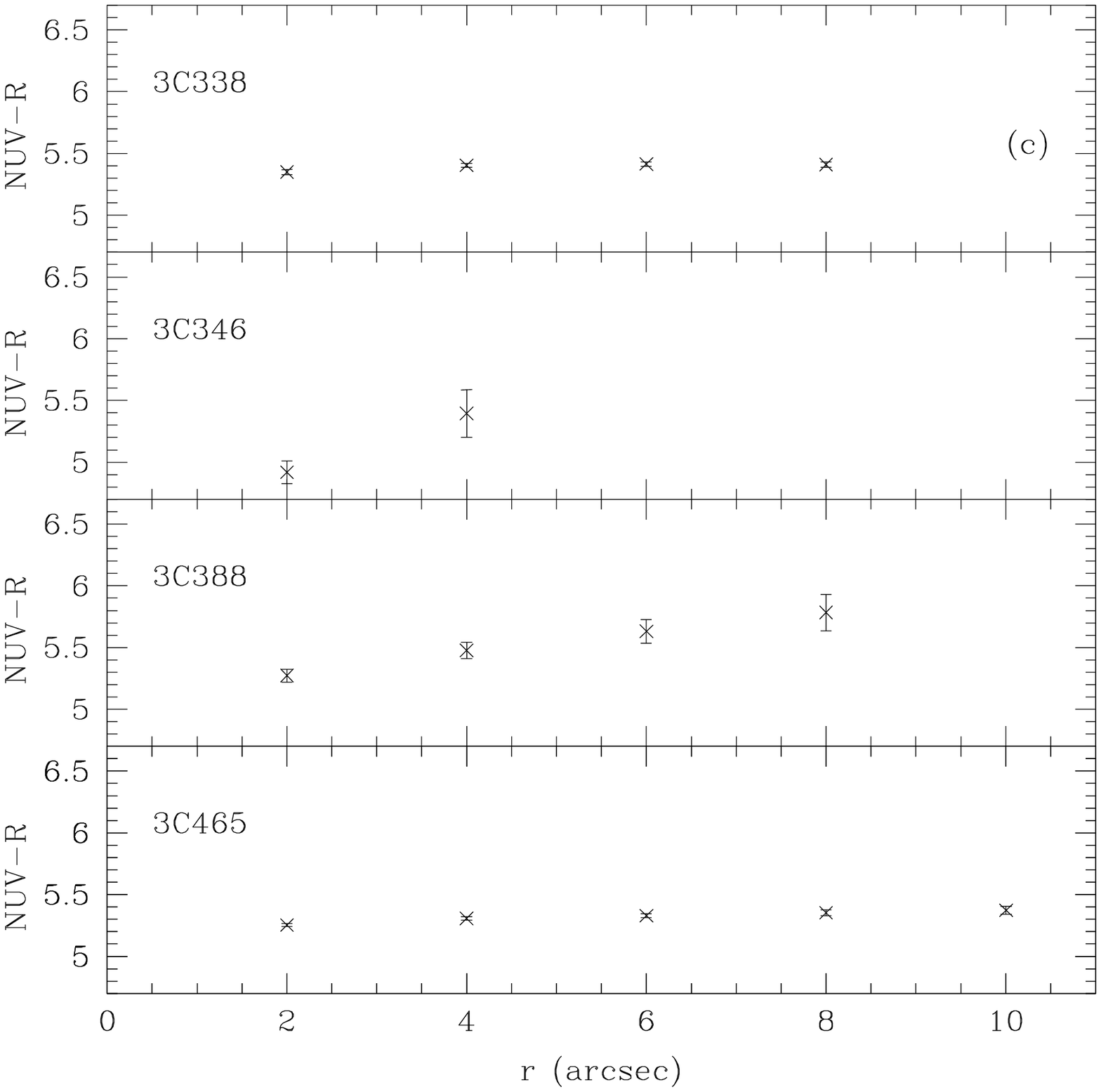,width=0.33\linewidth}}
\centerline{
\psfig{figure=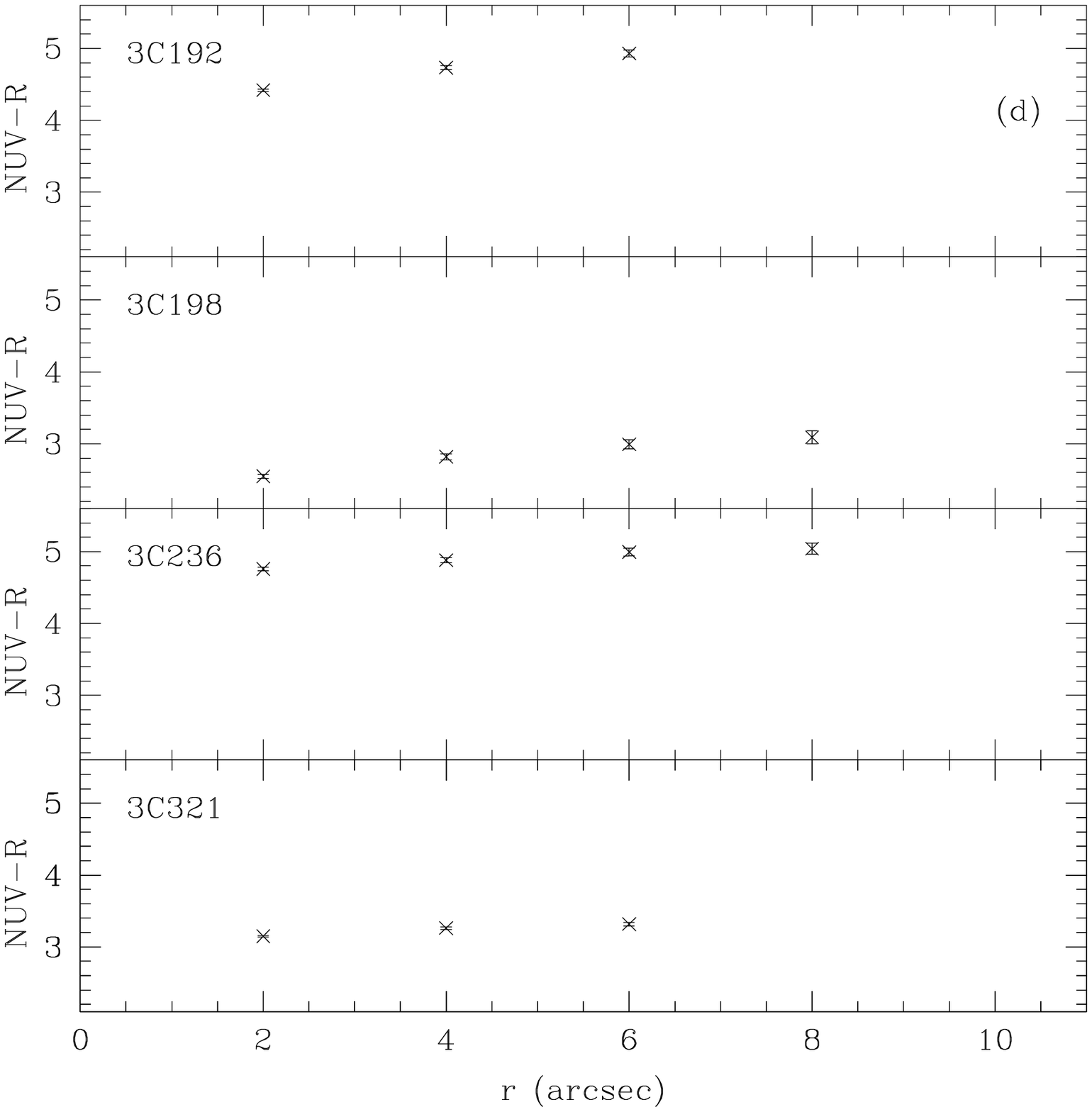,width=0.33\linewidth}
\psfig{figure=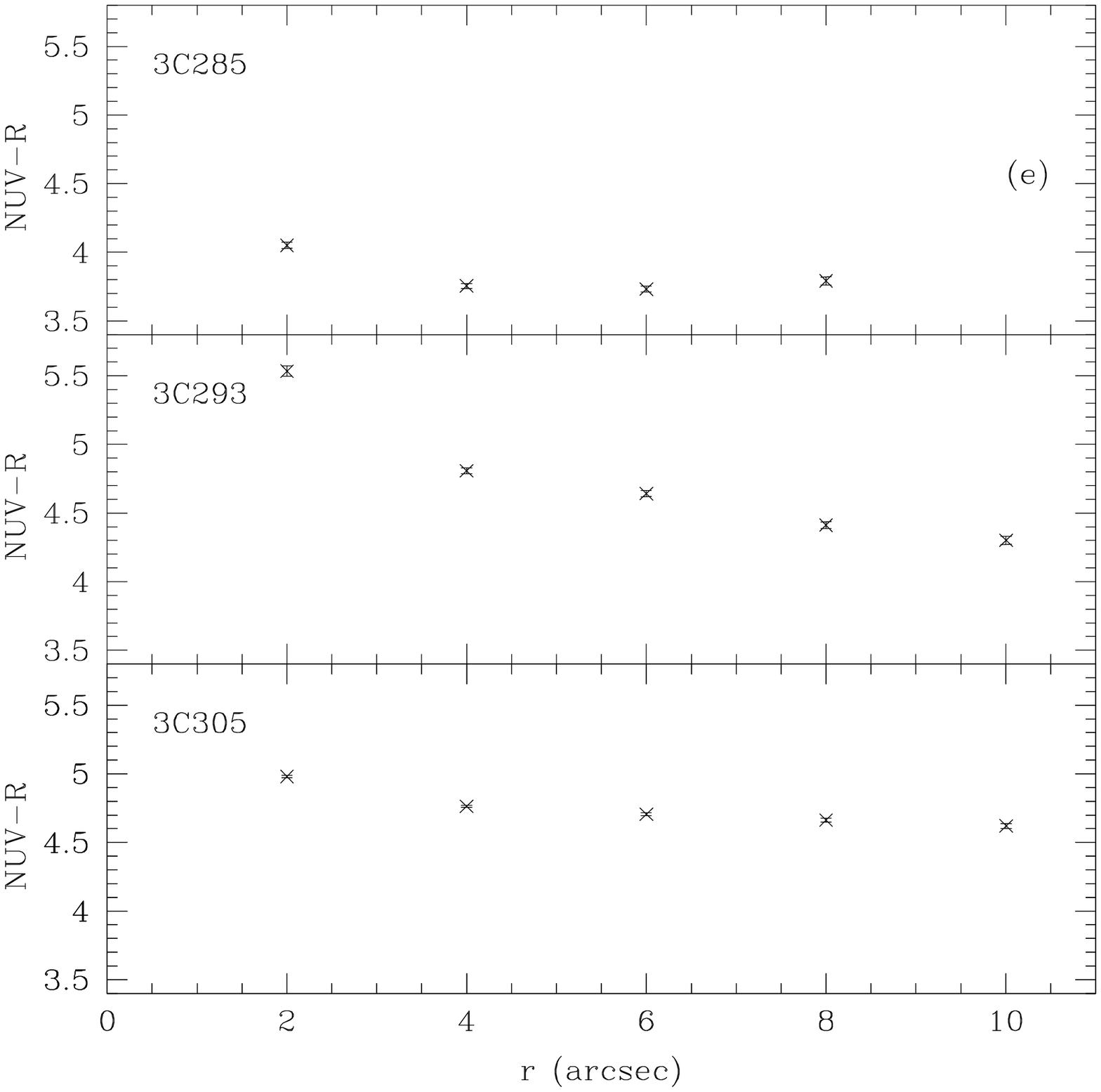,width=0.33\linewidth}
\psfig{figure=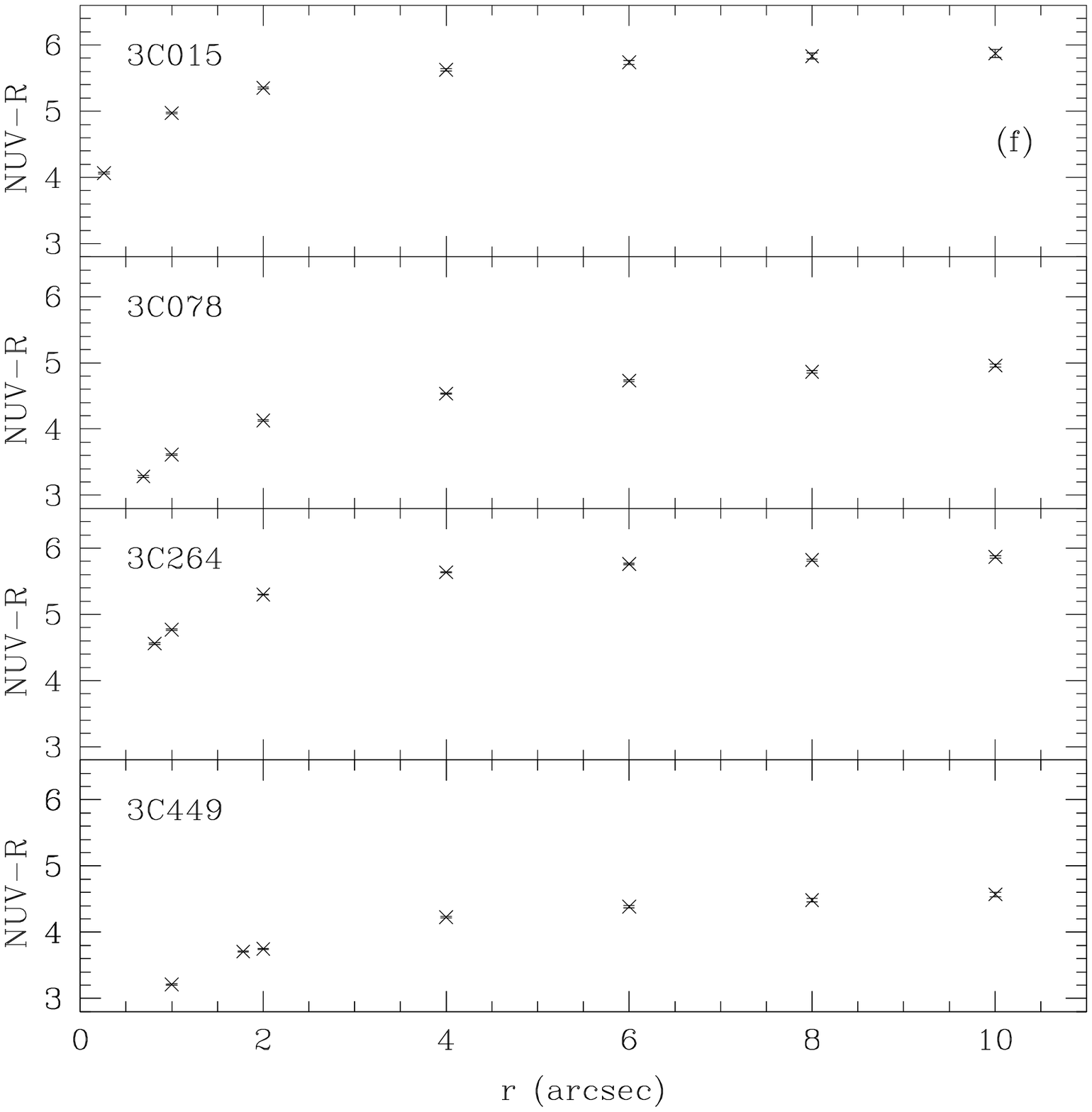,width=0.33\linewidth}}
\caption{The NUV-R color (HST based) profiles of the sample: (a),(b),(c) red
quiescent galaxies; (d), and (e) blue UV clumpy galaxies, separated into NUV-r
color increasing and decreasing with radius, (f) blue UV-disky galaxies. For
these last objects we present also the NUV-R color at 1$^{\prime\prime}$ and
at the aperture corresponding to their dusty disks in order to appreciate
better the color profiles bluing to smaller radii.}
\label{prof}
\end{figure*}

After these procedures, we obtained the NUV-R color profiles for 25 3CR
radio-galaxies, which are shown in Fig. \ref{prof} (grouped into the
categories described in Sect. \ref{results}). The NUV-R colors obtained were
corrected for Galactic absorption by an amount equal to 2.48$\times$E(B-V),
derived from the extinction in the form given by \citet{cardelli89}.  In most
cases (and we highlight the few exceptions below), the color profiles were
flat with color variations smaller than $\pm 0.2$ mag.  Therefore, the color
for each galaxy was almost independent of the aperture used.

\begin{table*}
\caption{Observation log}
\label{tab1}
\centering
\begin{tabular}{ l c c c c c c c}
\hline\hline
Name & \multicolumn{2}{c}{\,\, Filter \,\,\,  Exp. time} & Date & Opt CCC & UV CCC & Opt Jet & UV Jet\\
\hline	       			      	 	                  
3C~015   & F25QTZ   & 7230 & 01-01-27 &  YES    &  YES &  NO  &  YES \\
3C~029   & F25SRF2  & 1440 & 01-01-27 &  YES    &  YES &  NO  &  NO  \\
3C~035   & F25SRF2  & 1440 & 99-10-21 &  NO     &  NO  &  NO  &  NO \\
3C~040   & F25SRF2  & 1440 & 00-06-03 &  NO     &  NO  &  NO  &  NO \\
3C~066B  & F25SRF2  & 1440 & 99-07-13 &  YES    &  YES &  NO  &  YES \\
3C~078   & F25QTZ   & 2000 & 00-03-15 &  YES    &  YES &  YES &  YES \\
3C~192   & F25SRF2  & 1440 & 00-03-23 &  NO     &  NO  &  NO  &  NO \\
3C~198   & F25SRF2  & 1440 & 00-04-23 &  YES    &  YES &  NO  &  NO \\
3C~227   & F25SRF2  & 1440 & 00-01-25 &  YES    &  YES &  NO  &  NO  \\
3C~236   & F25SRF2  & 1440 & 99-10-05 &  NO     &  NO  &  NO  &  NO \\
3C~264   & F25QTZ   & 3600 & 00-02-12 &  YES    &  YES &  YES &  YES   \\
3C~270   & F25SRF2  & 1440 & 00-03-05 &  YES    &  NO  &  NO  &  NO  \\
3C~274   & F25QTZ   &  640 & 03-06-08 &  YES    &  YES &  YES &  YES \\
3C~285   & F25SRF2  & 1440 & 00-04-16 &  YES    &  NO  &  NO  &  NO \\
3C~293   & F25SRF2  & 1440 & 00-06-14 &  NO     &  NO  &  NO  &  NO   \\
3C~296   & F25SRF2  & 1440 & 00-04-15 &  YES    &  NO  &  NO  &  NO \\
3C~305   & F25SRF2  & 1440 & 00-04-27 &  NO     &  NO  &  NO  &  NO \\
3C~310   & F25SRF2  & 1440 & 00-06-10 &  YES    &  YES &  NO  &  NO   \\
3C~317   & F25SRF2  & 1440 & 99-07-26 &  YES    &  YES &  NO  &  NO   \\
3C~321   & F25SRF2  & 1440 & 00-06-05 &  NO     &  NO  &  NO  &  NO    \\
3C~326   & F25SRF2  & 1440 & 00-03-12 &  NO     &  NO  &  NO  &  NO   \\
3C~338   & F25SRF2  & 1440 & 00-06-04 &  YES    &  YES &  NO  &  NO    \\
3C~346   & F25QTZ   & 3600 & 00-08-22 &  YES    &  YES &  YES &  YES     \\
3C~353   & F25SRF2  & 1440 & 00-06-22 &  NO     &  NO  &  NO  &  NO    \\
3C~371   & F25QTZ   & 7989 & 00-09-21 &  YES    &  YES &  NO  &  NO    \\
3C~382   & F25CN182 & 1440 & 00-02-23 &  YES    &  YES &  NO  &  NO   \\
3C~388   & F25SRF2  & 1440 & 00-06-02 &  YES    &  YES &  NO  &  YES   \\
3C~390.3 & F25CN182 & 1440 & 99-08-10 &  YES    &  YES &  NO  &  NO    \\
3C~449   & F25SRF2  & 1440 & 00-04-16 &  YES    &  YES &  NO  &  NO  \\
3C~452   & F25CN182 & 1440 & 00-01-30 &  NO     &  NO  &  NO  &  NO   \\
3C~465   & F25SRF2  & 1440 & 00-05-25 &  YES    &  YES &  NO  &  NO  \\
\hline
\end{tabular}

Column description. Col. (1): 3CR source name. Col. (2): STIS filter
used. Col. (3): exposure time [s].  Col. (4): observation dates. Col. (5)-(8):
the presence of optical and UV central compact core (CCC) and jets in each
object as taken by \citet{chiaberge99} and \citet{chiaberge02}.
\end{table*}

\subsection{Cross-calibration of NUV-R colors}
\label{cross}

The previous studies of galaxy UV/optical color analysis were performed using
GALEX and SDSS observations, by measuring the NUV-r color, while we used NUV
and optical HST observations. Before we can compare our HST-based NUV-R color,
which for clarity we refer to as (NUV-R)$_{\rm {HST}}$, with results present
in the literature, we must complete a cross-calibration between GALEX, SDSS,
and HST data.

The GALEX near-ultraviolet filter is centered on $\lambda_{eff}$ = 2271 \AA
with $\Delta\lambda$ = 732 \AA. The images correspond to a circular field of
view of radius $\sim$38$\arcmin$ and a spatial resolution of $\sim
5\arcsec$. The Sloan Digital Sky Survey (SDSS) images (Fourth Data Release)
have a field of view of 13.51$\arcmin$$\times$8.98$\arcmin$ and a pixel size
of 0$\farcs$396. The r-filter band is centered at 6231 \AA\ with a FWHM of
1373 \AA.

\begin{figure}
\centerline{
\psfig{figure=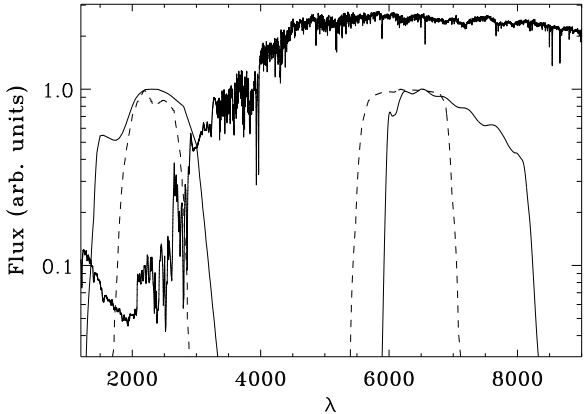,width=1\linewidth}}
\caption{The transmission curves of HST (STIS/F25SRF2 filter and WFPC2/F702W
  filter are marked with a solid line), GALEX-SDSS (NUV filter and r filter are
  marked with a dashed line) overplotted on a 
galaxy spectrum model taken from
  \citet{bc03} with 11 Gyr and Z = 0.008Z$_{\odot}$.}
\label{trasm}
\end{figure}

The filter transmission curves are shown in Fig. \ref{trasm}, with a typical
spectrum of an old (11 Gyr) stellar population. We note that the STIS
passbands are wider than the GALEX/NUV passband, while there is a small shift
between the centers of the SDSS and HST/F702W optical filters.

Ideally, cross-calibrations should be performed by directly comparing
images of the same objects taken with different telescopes in the band of
interest. However, the object with the highest value of total counts in
the GALEX NUV image in our sample is 3C~040, observed as part of the MIS
survey.  However, even for this object the flux ratio between HST and GALEX
(0.7 $\pm$ 0.2) has an associated error that it is too large for a meaningful
cross-correlation.

We therefore decided to compare the GALEX and HST data using 
stellar population synthesis models; this was similar to the method
described by \citet{proffitt06}. We used 18 models from \citet{bc03} in a grid of
different age (1,3,5,7,9, and 11 Gyr) and metallicity (0.008, 0.02 [=
Z$_{\odot}$], and 0.05). We convolved these spectra with the different
transmission curves in NUV and optical bands to compare the resulting
fluxes and estimate the color correction to be applied to the HST colors.

\begin{figure*}
\centerline{
\psfig{figure=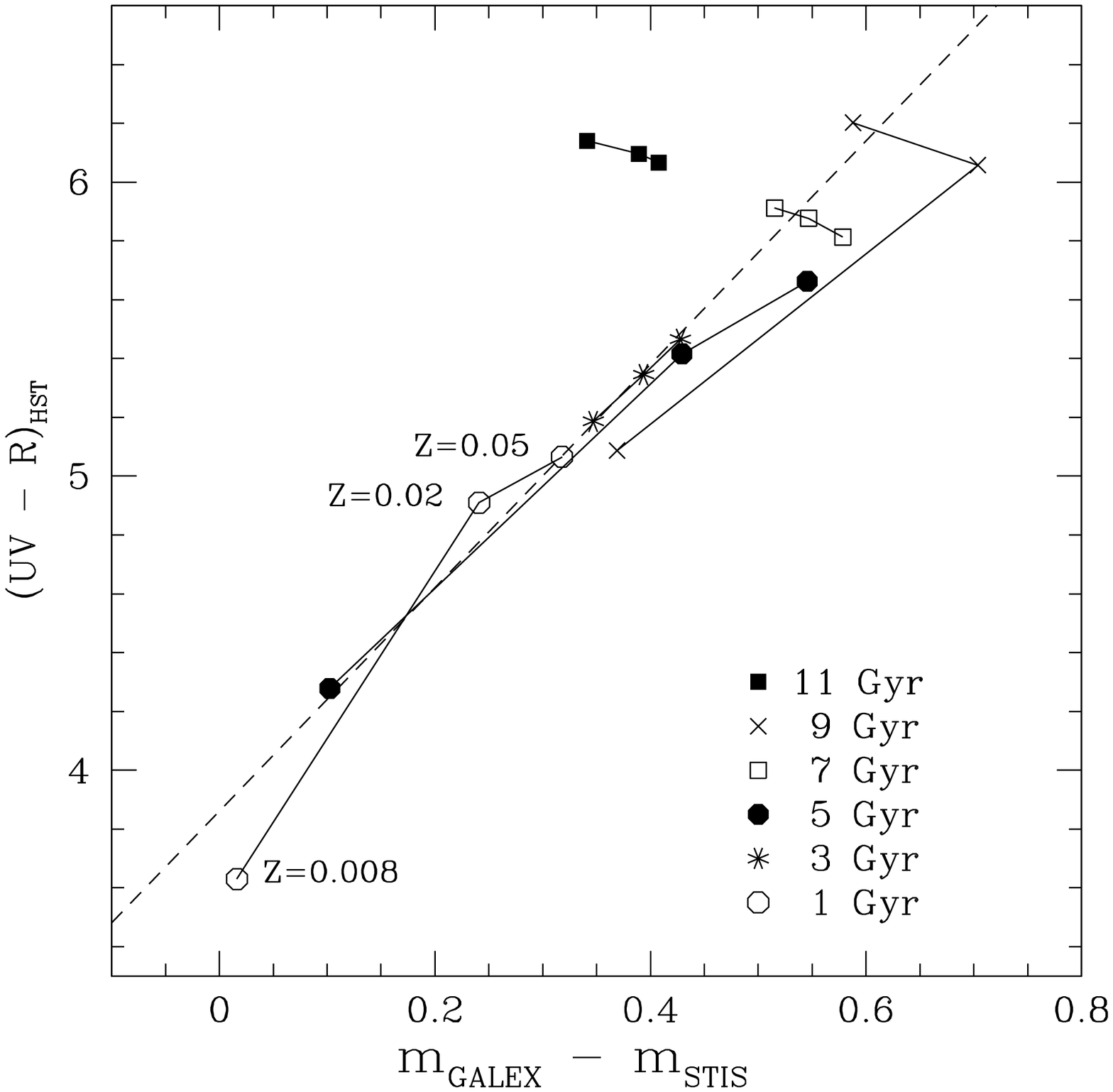,width=0.33\linewidth}
\psfig{figure=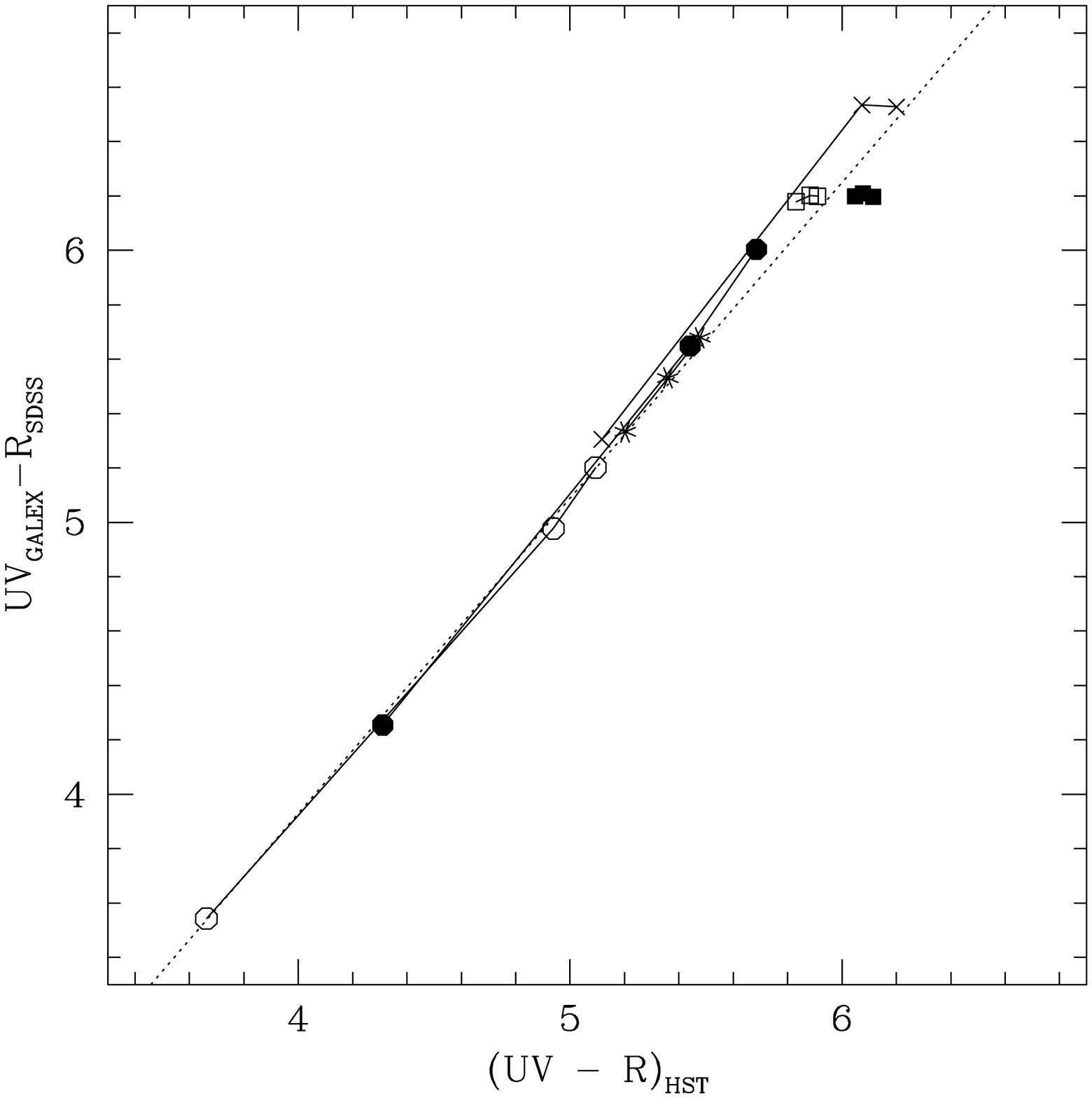,width=0.33\linewidth}
\psfig{figure=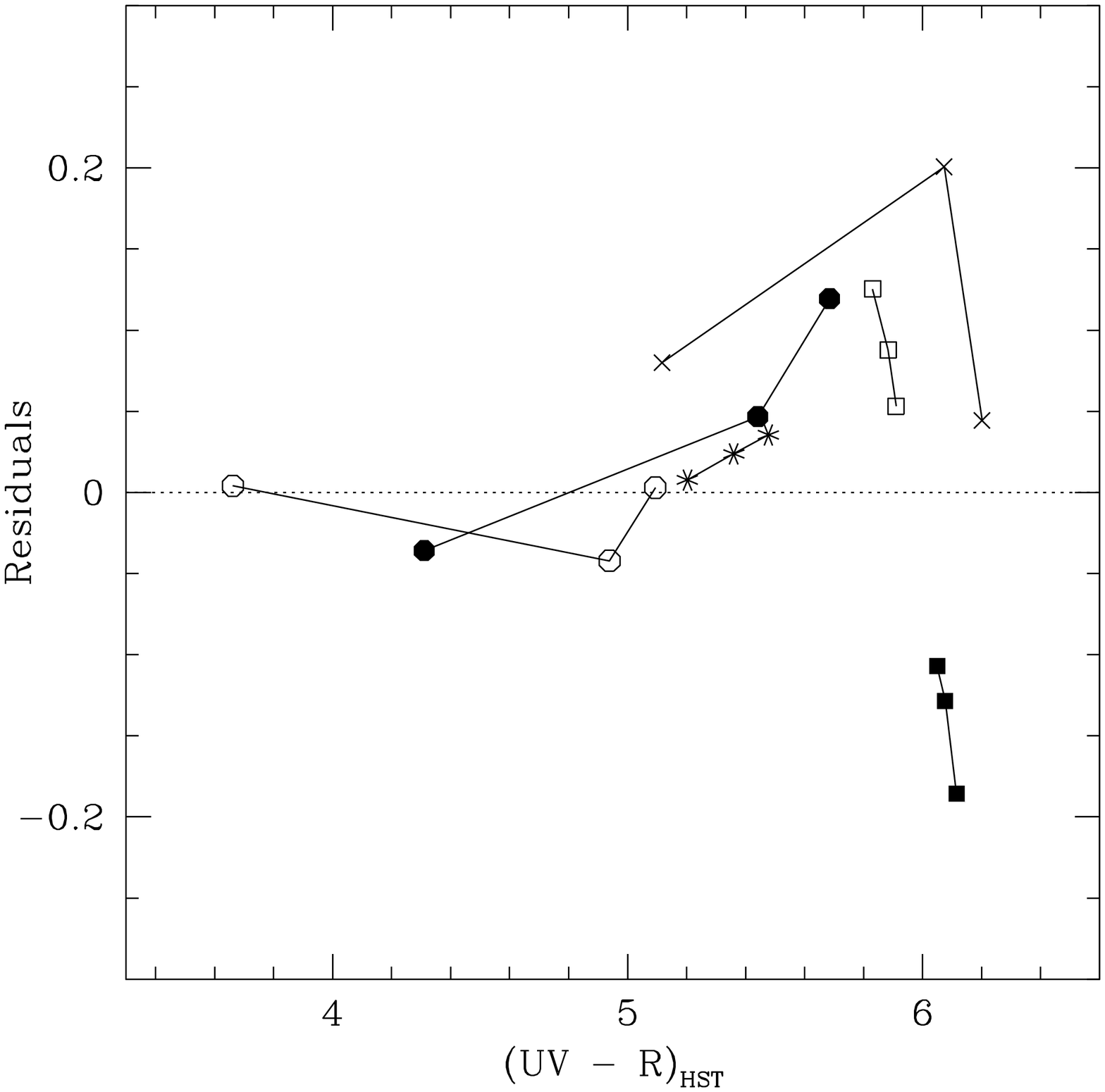,width=0.33\linewidth}}
\caption{Conversion from the HST NUV-R color to GALEX/SDSS NUV-r color,
estimated from models of stellar population synthesis from \citet{bc03} as
marked in the figure.  The left panel shows the linear relation between the
NUV magnitudes and the observed HST color.  The middle panel shows the
connection between HST and GALEX-SDSS color with overplotted the linear
relation we used for the color correction. The right panel shows the residuals
from this linear relation.}
\label{corr}
\end{figure*}

As shown in Fig. \ref{corr}, left panel, the conversion of the NUV magnitude
from STIS to GALEX has a significant dependence on the stellar population age
and metallicity. For a redder stellar population, the larger width of
the STIS filter towards longer wavelength, increases the measured STIS flux with
respect to the GALEX data.  The correction that has to be applied to the STIS
data has an almost linear dependence on (NUV-R)$_{\rm {HST}}$.

Conversely, the correction to convert the optical band images from the WFPC2
to SDSS bandpasses is far smaller and has only a weak trend with stellar
population, since in this band the spectrum is flat. In this case, we were
able to check directly the results of the stellar population analysis by
comparing the fluxes for the same aperture in the HST/WFPC2 (opportunely
convolved) and SDSS r-band images.  The results obtained confirmed the need
for only a small magnitude correction (a few hundredths of a magnitude).

In the middle panel of Figure \ref{corr}, we show the correspondence between the
NUV-R HST colors and the NUV-r color in the GALEX/SDSS system.  It can be
reproduced with a linear relation in the form:

\medskip
\noindent
(NUV-r) - (NUV-R)$_{\rm {HST}}$ = 0.16 $\times$ [(NUV-R)$_{\rm {HST}}$ - 4] - 0.07

\medskip

In the right panel of Figure \ref{corr}, we present the residuals from this
linear relation of $\sim$ 0.2 mag, which we adopt as a conservative error in
the color conversion.
     
We applied the same method used for the STIS/F25SRF2 filter to the STIS/F25QTZ
filter, which differs only slightly with short wavelength cutoff at longer
wavelength. The color conversion in this case is

\medskip
\noindent
(NUV-r) - (NUV-R)$_{\rm {HST}}$ = 0.15  $\times$ [(NUV-R)$_{\rm {HST}}$ - 4] - 0.04

\medskip

We also tested the cross-calibration between GALEX-SDSS and HST data by using
a two-stellar-populations galaxy model and varying the relative contribution
of a 1 Gyr and 11 Gyr stellar population models at three different
metallicities. We obtained a quasi-linear cross-calibration relation, which
links the points of the two individual populations in Fig. \ref{corr}. The
maximum residual is 0.2 mag, similar to our typical single-stellar-population
model cross-calibration error.

In Table \ref{tab2}, we indicate the aperture size used for each object and
the corresponding color.  This (NUV-R)$_{\rm {HST}}$ color was transformed
into the NUV- r color using the prescription described above.

\begin{table*}
\caption{Properties of the 3CR sample.}
\label{tab2}
\centering
\begin{tabular}{ l c c c c c c c c c c}
\hline\hline
3C      &      z     & log~$P_{178}$ & FR - E.L. & M$_{R}$ & E(B-V) & $\sigma$  & (NUV-R)$_{HST}$  & \multicolumn{2}{c}{aperture ($^{\prime\prime}$ \,\, kpc)} & UV morphology\\
\hline	
15      &   0.073    &    26.2 &    I/II - LEG & -22.41 & 0.022 &       & 5.87 $\pm$ 0.06$^*$   & 10 & 14.1 & disky  \\
29      &   0.0447   &    25.8 &         I     & -22.73 & 0.036 &  231  & 6.2  $\pm$ 0.1    & 10  & 7.0 &        \\
35      &   0.0670   &    26.0 &    II - LEG   & -22.34 & 0.141 &       & 5.6  $\pm$ 0.2    & 6  & 2.6  &   \\
40      &   0.0177   &    25.2 &    II - LEG   & -22.62 & 0.041 &  242  & 5.58 $\pm$ 0.02   & 10 & 3.5  &       \\
66B     &   0.0215   &    25.4 &         I     & -23.42 & 0.080 &       & 5.76 $\pm$ 0.03   & 8  & 3.3  &       \\
78      &   0.029    &    25.5 &         I     & -23.23 & 0.173 &  263  & 4.96 $\pm$ 0.03$^*$   & 10 & 5.6  & disky  \\
192     &   0.060    &    26.2 &    II - HEG   & -21.60 & 0.054 &       & 4.93 $\pm$ 0.05   & 6  & 6.9  & knotty  \\
198     &   0.0815   &    26.1 &    II - HEG   & -20.82 & 0.026 &  174  & 3.09 $\pm$ 0.09   & 8  & 12.6 & knotty   \\
227     &   0.0861   &    26.7 &    II - WQ    & -21.42 & 0.026 &       & nucleated         &    &      &       \\
236     &   0.0989   &    26.5 &    II - HEG   & -23.10 & 0.011 &       & 5.04 $\pm$ 0.08   & 8  & 11.7 & knotty   \\
264     &   0.0217   &    25.4 &         I     & -22.51 & 0.023 &  271  & 5.87 $\pm$ 0.02$^*$   & 10 & 4.2  & disky \\
270     &   0.0073   &    24.8 &         I     & -22.11 & 0.018 &  309  & 5.21 $\pm$ 0.01   & 8  & 1.2  &         \\
274     &   0.0037   &    25.6 &         I     & -22.39 & 0.022 &  333  & 5.14 $\pm$ 0.01   & 10 & 0.85 &          \\
285     &   0.0794   &    26.1 &    II - HEG   & -22.26 & 0.017 &  181  & 3.79 $\pm$ 0.03   & 8  & 12.3 & knotty  \\
293     &   0.0452   &    25.8 &    I/II - LEG & -22.10 & 0.017 &  201  & 4.30 $\pm$ 0.03   & 10 & 8.7  & knotty  \\
296     &   0.0237   &    25.2 &         I     & -23.38 & 0.025 &       & 5.59 $\pm$ 0.01   & 10 & 4.8  &         \\
305     &   0.0414   &    25.8 &    I/II - HEG & -22.88 & 0.029 &  193  & 4.62 $\pm$ 0.02   & 10 & 8.1  & knotty   \\
310     &   0.0540   &    26.5 &         I     & -22.30 & 0.042 &  209  & 5.4 $\pm$ 0.1     & 10 & 6.3  &       \\
317     &   0.0350   &    26.1 &         I     & -23.18 & 0.037 &  216  & 5.21 $\pm$ 0.04   & 10 & 6.7  &        \\
321     &   0.0960   &    26.4 &    II - HEG   & -22.58 & 0.044 &       & 3.32 $\pm$ 0.03   & 6  & 11.2 & knotty  \\
326     &   0.0885   &    26.5 &    II - LEG   & -21.71 & 0.053 &       & 6.0 $\pm$ 0.2     & 4  & 3.5  &        \\
338     &   0.0298   &    26.0 &         I     & -23.50 & 0.012 &  310  & 5.41 $\pm$ 0.02   & 8  & 4.7  &      \\
346     &   0.1610   &    26.8 &    II - LEG   & -22.11 & 0.067 &       & 5.4  $\pm$ 0.2    & 4  & 12.6 &     \\ 
353     &   0.0304   &    26.7 &    II - LEG   & -21.36 & 0.439 &       & 5.0  $\pm$ 0.5    & 2  & 1.2  &      \\
371     &   0.051    &    26.5 &    II - BLLAC & -22.66 & 0.036 &       & nucleated         &    &      &    \\
382     &   0.0578   &    26.2 &    II - WQ    & -23.50 & 0.070 &       & nucleated         &    &      &       \\
388     &   0.0908   &    26.6 &    II - LEG   & -23.07 & 0.080 &  408  & 5.8 $\pm$ 0.2     & 8  & 10.7 &         \\
390.3   &   0.0561   &    26.5 &    II - WQ    & -22.21 & 0.071 &       & nucleated         &    &      &       \\
449     &   0.0171   &    24.9 &         I     & -21.96 & 0.167 &  253  & 4.57 $\pm$ 0.03$^*$   & 10 & 3.3  & disky   \\
452     &   0.0811   &    26.9 &    II - HEG   & -22.21 & 0.137 &       &  6.0 $\pm$ 1.6    &  2 & 3.1  &       \\
465     &   0.0293   &    25.8 &         I     & -23.64 & 0.069 &  356  & 5.37 $\pm$ 0.03   & 10 & 5.9  &      \\
\hline
\end{tabular}

Column description. Col. (1): 3CR source name. Col. (2): redshift (from
NED). Col. (3): Log radio power (W Hz$^{-1}$) at 178 MHz taken by
\citet{allen02} and \citet{kellermann69}. Col (4): radio morphological and
optical spectral classification, as taken from \citet{jackson97} and from the
website http://www.science.uottawa.ca/$^{\sim}$cwillott/3crr/3crr.html with
the only variation from the literature is the classification of 3C~236 in HEG
derived from SDSS data. Col. (5): magnitude in R band
as taken from \citet{donzelli07} and from NED. Col. (6): galactic extinction
(from NED). Col. (7): stellar velocity dispersion taken from Hyperleda, the
website http://www.mpa-garching.mpg.de/SDSS/DR4/raw\_data.html and
\citet{smith90}. Cols. (8): NUV-r integrated color on the greater available
aperture with the corresponding scale radius in (9) arcsec and (10) kpc. The
objects labelled with * are the UV-disky galaxies, whose (NUV-R)$_{HST}$
colors, corrected excluding the UV disks contaminated by the emission
lines (see Sect. \ref{origin}), are: 6.1 $\pm$ 0.1 for 3C~015 , 5.24 $\pm$ 0.08
for 3C~078, 6.06 $\pm$ 0.06 for 3C~264, and 5.2 $\pm$ 0.1 for 3C~449. These
NUV-r colors are those that we finally use in the plots. Col. (11): UV
morphology.
\end{table*}

\section{RESULTS}
\label{results}

As described by \citet{schawinski06}, the NUV-r color can be used as a means
of identifying recent star formation.  They adopted a threshold at NUV-r $<$
5.4 to separate ``red'' quiescent galaxies from ``blue'' galaxies with active
star formation.  This value is suggested by both theoretical population
synthesis models, although the availability of these models is limited, and
empirical data \citep{lee05}. Despite the possible presence of an UV upturn,
observed often in early-type galaxies (\citealt{code79} and
\citealt{burstein88}), passively evolving galaxies do not appear to have NUV-r
colors bluer than 5.4. Given our 0.2 mag calibration uncertainty, we consider
only galaxies bluer than NUV-r $<$ 5.2 to be bona-fide ``blue'' galaxies.

\begin{figure*}
\centerline{
\psfig{figure=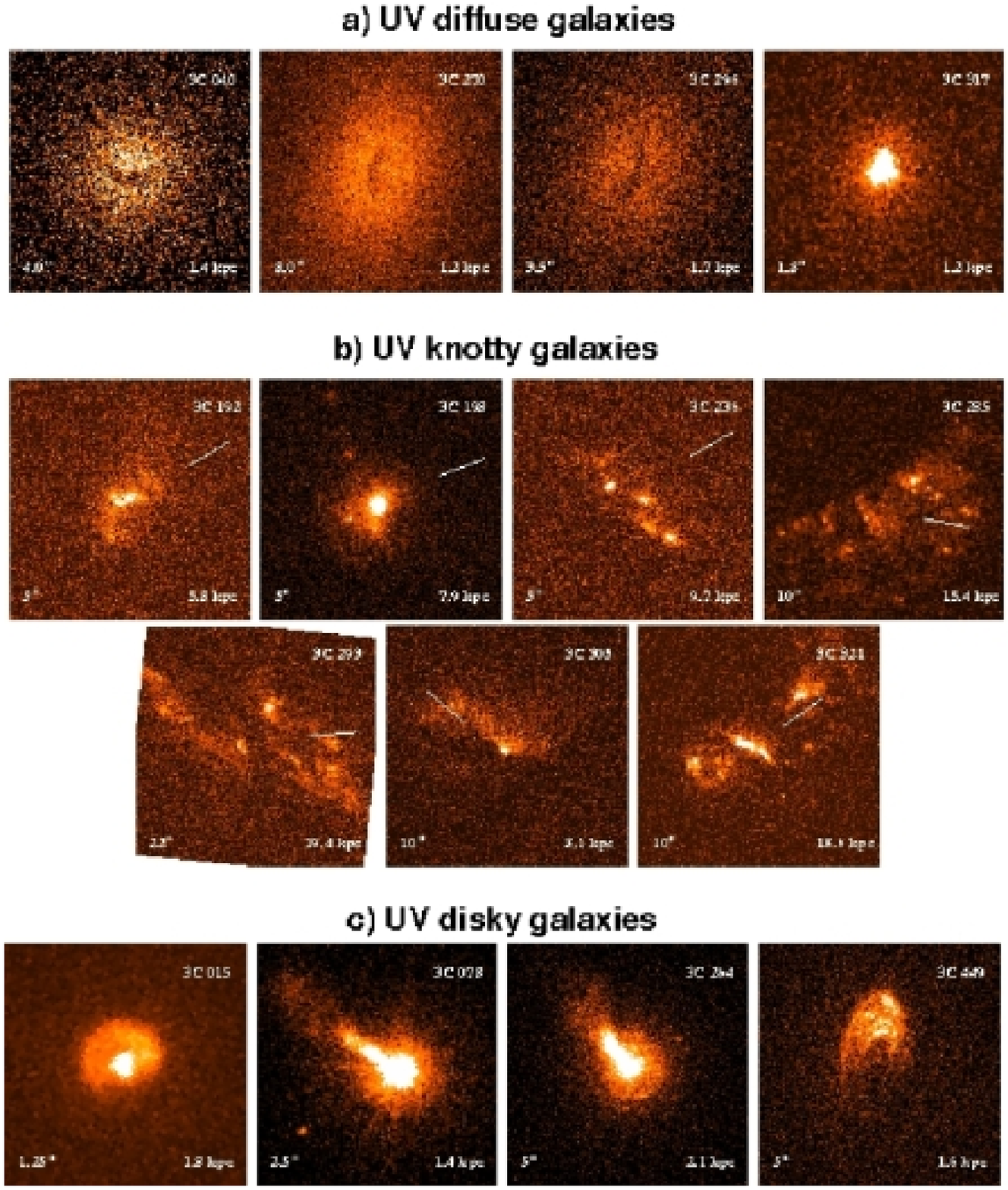,width=0.9\linewidth}}
\caption{a) The UV images of 4 illustrative examples of red diffuse
  galaxies. Note the UV nucleus of 3C~317, masked in our analysis, superposed
  to the diffuse emission. b) The UV images of the blue UV-clumpy galaxies
  show present star formation on large scale ($\sim$5-20 kpc) arranged in
  discrete knots. We also show the orientation of their radio-axis. c) Images
  of the UV-disky galaxies which show, apart from their UV jets and nuclei,
  UV emission in a small scale circumnuclear disks ($\sim$0.5-1 kpc).}
\label{immgal}
\end{figure*}

On the basis of this criteria for the integrated colors and UV morphology, the
galaxies of the sample can be divided in 3 main categories:

\begin{enumerate}
\item{quiescent red galaxies}: 14 object have red and flat integrated color
profiles (Fig. \ref{prof} a,b,c).  They lack any sign of significant recent
star formation, since, apart from emission from the active nucleus (and from
extended UV/optical jets), we detect only diffuse UV light tracing the optical
emission (Fig. \ref{immgal}, a panel). We note that this description applies
to the 7 galaxies with 5.2 $<$ NUV-r $<$ 5.6, whose classification is
uncertain due to our calibration accuracy of 0.2 mag.

\item{blue UV-clumpy galaxies}: 7 objects have NUV-r $<$ 5.2 and show a clumpy
UV morphology (and they are labeled as ``knotty'' in Table \ref{tab2}), which
are usually associated with dust lane structures (see Fig. \ref{immgal}, b
panel). Their color profiles show different trends, since in some cases NUV-r
increases toward larger radii (Fig. \ref{prof}, d panel), but the opposite
behavior is also observed (e panel). Their UV emission is extended over 5-20
kpc.

\item{UV-disky galaxies}: 2 objects (namely 3C~078 and 3C~449) have a blue
NUV-r color, but the UV emission is located co-spatially with the
circum-nuclear dusty disk (Fig. \ref{immgal}, c panel). Another 2 galaxies
(namely 3C~015 and 3C~264), although globally red, show similar UV emission,
which is associated with their disks. In all cases, the NUV-r color increases toward larger
radii and, by extending the color profile to apertures smaller than
2$^{\prime\prime}$ in 3C~015 and 3C~264, they also reach a blue color. All
of these bright UV disks have scales of 0.5-1 kpc and these galaxies are labeled
as disky in Table \ref{tab2}.
\end{enumerate}

\subsection{Origin of UV excess}
\label{origin}

As explained above, in quiescent early-type galaxies a color bluer than NUV-r
$<$ 5.4 can be considered as evidence for the presence of a young stellar
populations. However, in active galaxies, it is fundamental to assess the
origin of this UV emission. Other emission processes related to the presence
of an AGN can contribute to this excess, in particular UV nuclear scattered
light, nebular continuum, and emission lines. Emission line images represent a
useful tool to tackle the issue of AGN UV contamination. The sites where
emission lines are produced mark the geometrical intersection between the
interstellar medium and the nuclear ionizing light, i.e. the regions where
both UV lines and UV scattered light can be found. This analysis is crucial in
particular for the ``blue'' objects.

\begin{figure}
\centerline{
\psfig{figure=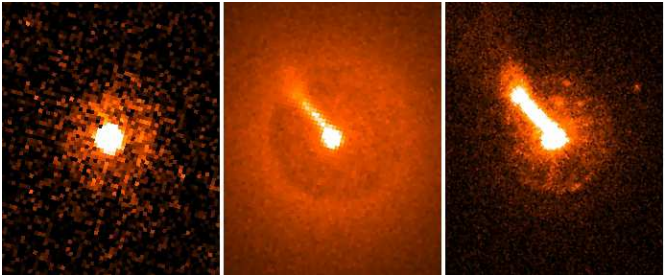,width=1\linewidth}}
\caption{The images (from the left to the right panel) of 3C~264 in
  H$\alpha$, in optical band (HST-F702W) and in NUV band (HST-F25QTZ). We
  note the spatial correspondence between the H$\alpha$ and UV emission
  region with the dusty disk.}
\label{riga3c264}
\end{figure}

We focused on the UV-disky galaxies with UV emission, that is cospatial with a
circum-nuclear dusty disk. We then considered H$\alpha$ images from the HST
archive, which are available for 3C~078, 3C~264, and 3C~449. In all 3
galaxies, there was a clear morphological correspondence between the UV and
H$\alpha$ emitting regions (for example see Fig. \ref{riga3c264}), which
suggests a possible contribution from UV emission lines. We then quantified
the fraction of contamination of emission lines to the total UV light by
opportunely scaling the H$\alpha$ flux.  We adopted the line ratios measured
by \citet{dopita97} in the emission line disk of M87, a well studied analogue
to our disked sources, for which HST UV spectroscopy is available.  We found
that UV emission lines contribute for between 50 and 80 \% of the UV light
within the area covered by the disk, with the strongest contribution
originatin in C IV$\lambda 1550$. The contamination would be even stronger
adopting the UV/optical line ratios measured by \citet{ferland86} in the NLR
region of nearby AGN, where the C~IV line is more prominent. These
results (that we extend for analogy to the fourth galaxy in this group,
3C~015, already classified as a red object) suggest that the disk-like
structures seen in UV light are due to line emission and not to a young
stellar population. Excluding the disk regions, both ``blue'' UV-disky objects
have a redder integrated NUV-r color of 5.3 and 5.4, for 3C~449 and 3C~078
respectively.

\begin{figure*}
\centerline{
\psfig{figure=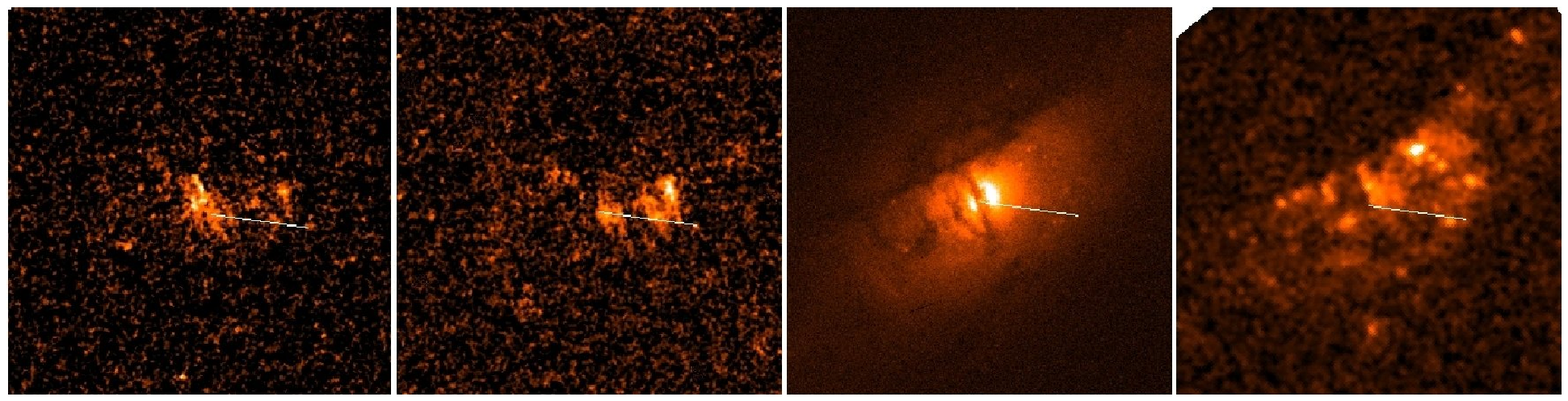,width=1\linewidth}}
\centerline{
\psfig{figure=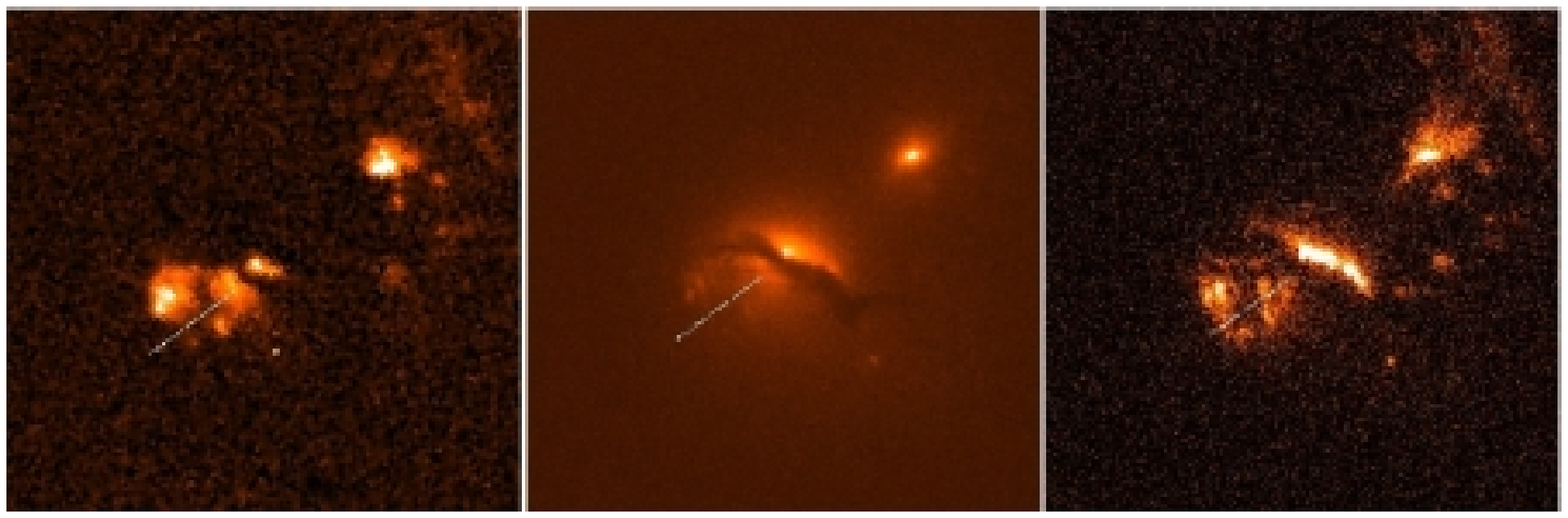,width=1\linewidth}}
\centerline{
\psfig{figure=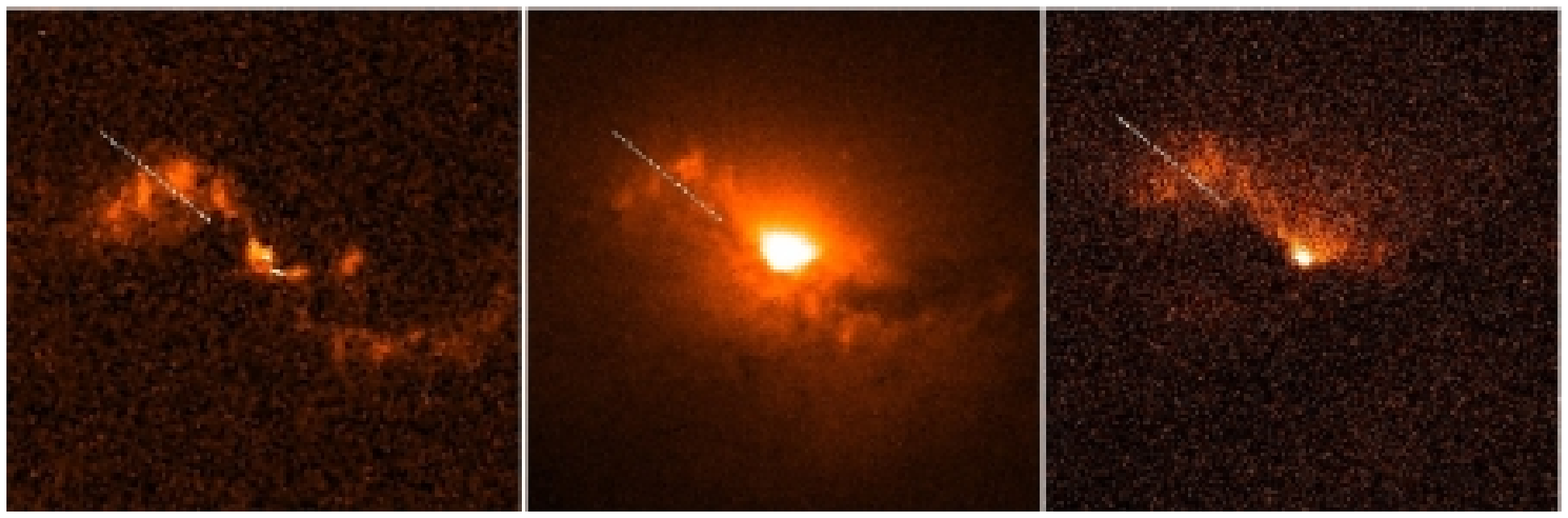,width=1\linewidth}}
\caption{The images (from the upper to lower panel) of 3C~285, 3C~321, and
  3C~305 in (from the left to the right panel) H$\alpha$ and [O III]
  (available only for 3C~285), in optical band and in NUV band. The size
    of all images are 10$\arcsec$ $\times$ 10$\arcsec$, corresponding to
    physical scales of 15.4, 18.6, and 8.1 kpc for 3C~285, 3C~321, and 3C~305
    respectively. We also show the orientation of their radio-axis.}
\label{rigaHEG}
\end{figure*}

We then focused on galaxies that present a clumpy morphology in UV
emission. In analogy with the analysis performed on UV-disky galaxies, we
selected emission-line images (in H$\alpha$ and [O~III]) from the HST archive,
which were available for 3C~285, 3C~305, and 3C~321. We then considered each
object individually: 1) in 3C~285, the H$\alpha$ emission was cospatial with
the dust feature bisecting its nuclear region, while the [O~III] emission is
confined to a one-sided elongated structure well aligned with its radio axis
(see Fig. \ref{rigaHEG} top panels).  There appears to be no significant
contribution by AGN related processes to the large scale UV emission in this
galaxy, since this is far more extended and not cospatial with the
line-emitting regions; 2) in 3C~321 (reminding that the companion galaxy on
the NW side was masked in our analysis), the H$\alpha$-line image presents a
well defined conical morphology centered on the SE radio jet. The same
structure is seen in UV light and this provides strong support for AGN
contamination (either scattered light or line emission) in this region.
However the UV light in this object originates for about 2/3 of the total
flux from the chain of knots associated with the nuclear dust lane that
most likely represents an extended region of star formation, since it has no
counterpart in the H$\alpha$ image and is unrelated to the radio-structure; 3)
in 3C~305, the emission line takes the form a S-shaped structure, aligned with
the radio axis, reminiscent of what is often found in Seyfert galaxies
\citep[e.g. Mrk 3,][]{capetti:mrk3}.  The UV emission has a similar structure
and this indicates possible AGN contamination\footnote{As discussed below not
from nebular processes, but eventually from scattered light.}. However, the
location of the UV emission knots, cospatial with the nuclear dust structure,
might support the possiblity that they are star-forming regions. Based on our
data alone, we cannot reach a robust conclusion. However, stellar population
synthesis models applied to long slit spectra of 3C~305 imply that a young,
spatially extended, stellar population is present\citep{tadhunter05} and we
therefore consider this object to be an actively star-forming galaxy.

For the remaining 4 objects, whose HST emission-line images are unavailable,
we note that, in the case of scattered nuclear UV light, as expected by the
AGN unification model, the UV light should be aligned with the radio-axis
as observed for 3C~285 and 3C~321. 
Although such UV/radio alignment might possibly be present in 3C~192, 
no association between radio and UV light
in the other 3 UV-knotty galaxies is present. 
A stronger link is instead present
with the dust structures, as already noted by \citet{allen02} from the
inspection of the optical HST images.  For example, in both 3C~305 and 3C~321
the brighter UV knots closely follow the edges of the dust lanes bisecting
their host galaxies. A similar association is seen also in 3C~285 and 3C~293.

We now consider the possibility of an emission line origin for the UV excess.
All seven ``blue'' UV-clumpy galaxies were observed with the F25SFR2 filter,
including Ly$\alpha$ at the edge of the transmission curve. Given their
redshift, this corresponds to a low transmission efficiency (0.02-0.07
relative to the transmission peak).  With a filter width of about 1000 \AA,
significant contamination is expected only for extreme values of the
Ly$\alpha$ equivalent width of approximately $\sim$15000-50000 \AA.  More
quantitatively, the Ly$\alpha$ fluxes can be measured from the H$\beta$ fluxes
(see Table \ref{contam}) taken from our own optical spectroscopy
\citep{butti08}, by adopting a scaling of F$_{{\rm Ly}\alpha}$/F$_{{\rm
H}\beta}$ = 55 \citep{ferland86}.  Line contamination is estimated to be
typically 1 - 4 \% and has a negligible effect on the integrated colors.
Furthermore, since the line emission tends to be strongly nuclear concentrated
\citep{mulchaey96}, most is contained within the nuclear region and is
therefore excluded from our UV flux measurements. Concerning the other bright
UV lines (i.e. C IV$\lambda 1549$, C III]$\lambda 1909$, and Mg II$\lambda
2798$), they are a factor of 5 to 30 times fainter than Ly$\alpha$; despite
the higher throughput of the F25SFR2 filter at their respective wavelengths,
their contribution to the UV flux is therefore also negligible.

Nebular continuum emission also contributes to the UV emission.  We estimated
that the UV contamination from nebular continuum at $\sim$ 2300 \AA\ (the
center of the passband of the STIS images) is 0.004 $\AA^{-1}$ H$\beta$ in the
low density limit and for a temperature of 10$^4$ K, adopting the coefficients
given by \citet{osterbrock89}. The contamination from nebular continuum is
estimated on a source-by-source basis (from the values reported in Table
\ref{contam}). In a similar way to the UV emission-line contamination, the
nebular continuum affects the global NUV-r color by less than approximately
0.1 mag (see the percentages of nebular contamination in Table \ref{contam}).

\begin{table}
\caption{Contribution of the nebular continuum in the blue UV knotty galaxies}
\label{contam}
{\centering
\begin{tabular}{ l c c c c }
\hline\hline
Name     & UV cont.&   H$\beta$    & neb. cont. (\%)  &  neb. cont (\%)\\
         &         &               & photometric  &  spectroscopic \\
\hline	    	                 
3C~192   &   10 &  32 & 11 & 9.7      (1) \\
3C~198   &   22 &  34 &  6 & $<$1     (2) \\
3C~236   &   10 &  13 &  5 & $<$1     (2) \\
3C~285   &   24 &   6 &  1 & 1.8      (2) \\
3C~293   &   46 &   6 &  1 & $<$10    (3) \\
3C~305   &   68 &  26 &  2 & $<$10    (3) \\
3C~321   &   33 &   4 &  1 & 0.3-25.2 (2) \\
\hline
\end{tabular}}

\medskip
Column description: (1): source name; (2): STIS UV flux in $10^{-17}\ergscmA$
units; (3) H$\beta$ fluxes in $10^{-16}\ergscm$ units; (4) percentage of UV
contamination from nebular continuum with our photometric analysis and (5) with
spectroscopic analysis from literature (6): 1, \citet{wills02};
2, \citet{holt07}; 3, \citealt{tadhunter05} (the range depends on the size of
the aperture used).
\end{table}

Until now, we have not taken into account the effects of internal reddening,
which is clearly an important issue considering the large $\lambda$ coverage
of the NUV-R color and the presence of dust in several targets. We discuss
the reddening effects for our sample, considering separately the three
different classes of galaxies:

\begin{enumerate}
\item{knotty blue galaxies: they all indeed have extended dust
    structures. However, the reddening-corrected NUV-R color would be even
    bluer than the uncorrected values. The presence of dust makes us
    overestimate the contamination from UV line emission and nebular
    continuum, strengthening the case for a young stellar population in these
    objects.}
\item{Red quiescent galaxies: they typically have no significant dust lanes
    that can absorb UV and optical light. This is confirmed by the flatness of
    their NUV-R color profiles at different radii. Therefore, the internal
    reddening effect in these objects is likely to be negligible.}
\item{Blue/red disk galaxies: the correction for internal reddening associated
    with their dusty disks would indeed make these objects bluer, and,
    furthermore, the absorption from this disk leads to an overestimate of the
    contamination from UV line emission, which we call their UV excess. We
    estimated the typical absorption of these disks, taking from the
    literature the V-R and R-I color excesses that they cause
    \citep{2000ApJS..130..267M}. The corresponding color excess is E(UV-R) =
    0.15 - 0.52. Therefore the line contamination is not significantly altered
    by absorption within these disks (since the line flux is reduced only by
    15-40\%) and it is sufficient to cause their blue colors, even after
    reddening correction.}
\end{enumerate}

We conclude that internal reddening causes to appear the knotty galaxies even
bluer and is negligible for the other objects. Therefore it does not affect
our main results.

All of these arguments imply that only a young stellar population can explain
the UV excess in these early-type galaxies, apart from the possible
exceptions of 3C~192.

\section{Discussion}
\label{discussion}

We focus on the global integrated NUV-r color of the objects in the 3C
sample. For the UV-disky galaxies, the NUV-r global color is measured from the
region of the galaxies outside their UV disks, contaminated by emission lines.

\begin{figure}
\centerline{
\psfig{figure=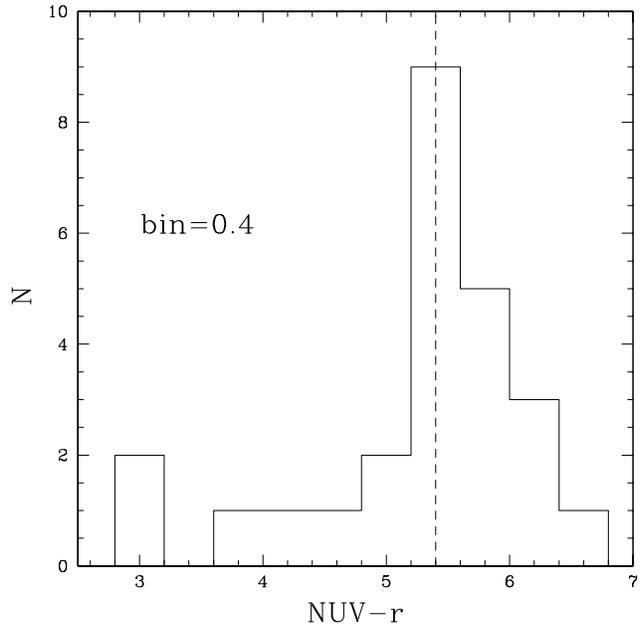,width=1\linewidth}}
\caption{The NUV-r color distribution of the 3CR host galaxies. This
distribution is peaked at NUV-r $\sim$ 5.4 but with a substantial blue tail.}
\label{isto}
\end{figure}

The NUV-r color distribution of our sample (see Fig. \ref{isto}) is peaked at
NUV-r $\sim$ 5.4 but with a substantial blue tail. Comparing our results with
those found by \citet{kauffmann06} for a volume-limited sample of AGN hosted
by massive bulge-dominated galaxies, we note that the color distribution of
their AGN, limited to those for the same range of stellar velocity dispersion
$\sigma$ ($\sigma$ $>$ 180 km/s), is similar to that estimated from our
sample.

\medskip

\begin{figure}
\centerline{
\psfig{figure=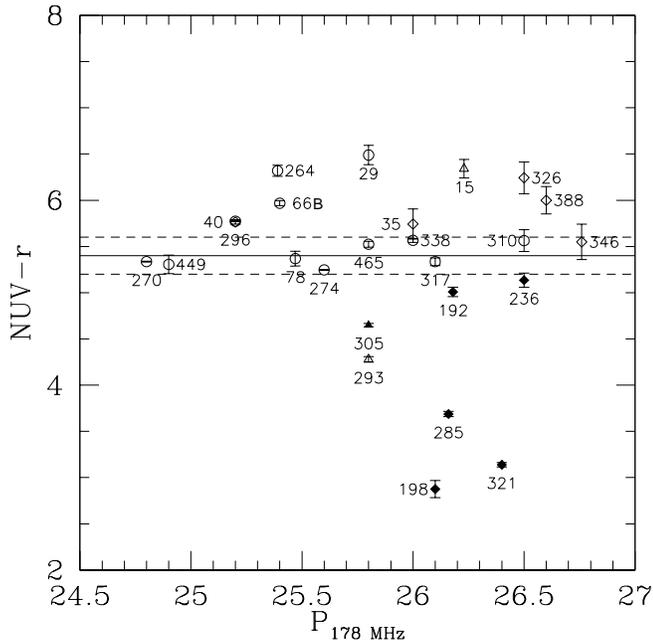,width=1\linewidth}}
\caption{NUV-r color versus radio power (178 MHz) taken by \citet{allen02} and
  \citet{kellermann69} for the sample. The solid line, drawn at NUV-r = 5.4,
  marks the separation between blue and red galaxies.  The two dashed lines
  represent the level of accuracy of the color calibration. The circles 
  represent the FR~I objects, the triangles the intermediate FR~I/II galaxies,
  and the diamonds FR~II galaxies. The empty points correspond to low
  excitation galaxies (LEG), while the filled ones to high excitation galaxies
  (HEG).}
\label{Lradio}
\end{figure}

We consider in more detail the connection between AGN and recent star
formation. First of all, the NUV-r color is not simply related to the radio
power (see Fig. \ref{Lradio}). While the bluest galaxies are all at relatively
high values of radio power at 178 MHz (P$_{178 \rm MHz}$), we also find many
of the reddest galaxies at the high end of radio luminosities.

\begin{figure}
\centerline{
\psfig{figure=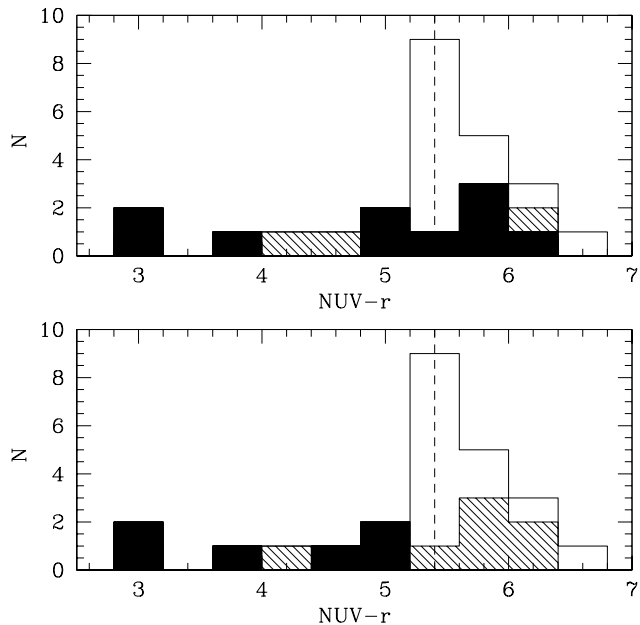,width=1\linewidth}}
\caption{The NUV-r color distribution of 3CR galaxies. In the upper panel we
point out the FR-morphological separation: the FR~II distribution is black,
the FR~I one is empty and the FR~I/II intermediate one is shaded. In the lower
panel we point out the excitation level classification: the HEG distribution
is black, the FR~II-LEG one is shaded and the distribution of the FR~I objects
is empty (for simplicity we sorted the FR~I/II intermediate-LEG objects as
FR~II-LEG).}
\label{isto2}
\end{figure}

In Fig. \ref{isto2} (upper panel), we separate the objects depending on their
FR morphological class \citep{fanaroff74}. All bona-fide blue galaxies belong
to the FR~II class or intermediate FR~I/FR~II objects, while no blue galaxy of
FR~I class is found. However, in the FR~II classes, we find almost equal
numbers of ``blue'' and ``red'' galaxies. No simple trend between NUV-r color
and radio morphology is therefore present.

We now investigate whether there is a connection between the extent of the
radio emission and color. At least in the case of FR~II, the radio-source size
is related to its age. A small radio source size could relate to a recent
onset of activity, triggered by a more recent merger, with a higher
possibility of detecting the induced event of star formation with respect to 
large radio sources. On the basis of these considerations, we measured
the radio source sizes of the FR~II galaxies from images in the literature.
For the blue galaxies, two objects, 3C293 and 3C305, were associated with a
small-scale, 4 and 5 kpc, respectively, radio sources.  All of the other 5
blue objects were instead several hundreds of kiloparsec in size. They
included also 3C~236, which \citet{odea01} suggested was just re-ignited
radio-source, based on the presence of a small-scale radio emission
superimposed on a large-scale emission.  For the red galaxies, 3C~353 and
3C~388 have small-scale radio emission (12 and 8 kpc, respectively), while the
remaining sources have far larger dimensions. There appears to be no simple
relation between the color and the radio emission size for FR~II galaxies.

Instead, if we separate the sample on the basis of their excitation level
using the diagnostic optical line ratios (i.e. into low and high excitation
galaxies, LEG and HEG, \citealt{laing94} and \citealt{jackson97}), the
situation is far clearer (see Fig. \ref{isto2}, lower panel). All six
high-excitation galaxies are blue and, apart from possibly 3C~192, they all
show evidence of recent star formation; the sub-sample of HEG is in fact
essentially coincident with the blue UV-clumpy sub-sample.  The fraction of
star-forming HEG is far in excess of what found by \citet{schawinski06} in
their sample of quiescent galaxies, which is 20-25 \% for objects in the same
range of velocity dispersion or galaxy luminosity.

Conversely, there is only 1 blue LEG (namely 3C~293) out of 19, including all
FR~I in this class. The substantial tail of blue objects found
by \citeauthor{schawinski06} (their Fig. 1) does not appear to be present in this class of
objects, and this strongly disfavours the possibility of enhanced star
formation in LEG hosts, although the possible effects of the different
environment should also be considered (see the discussion below).

Therefore, it appears that a connection between recent star
formation and nuclear activity in 3CR hosts is only 
present in high excitation galaxies.

\begin{figure}
\centerline{
\psfig{figure=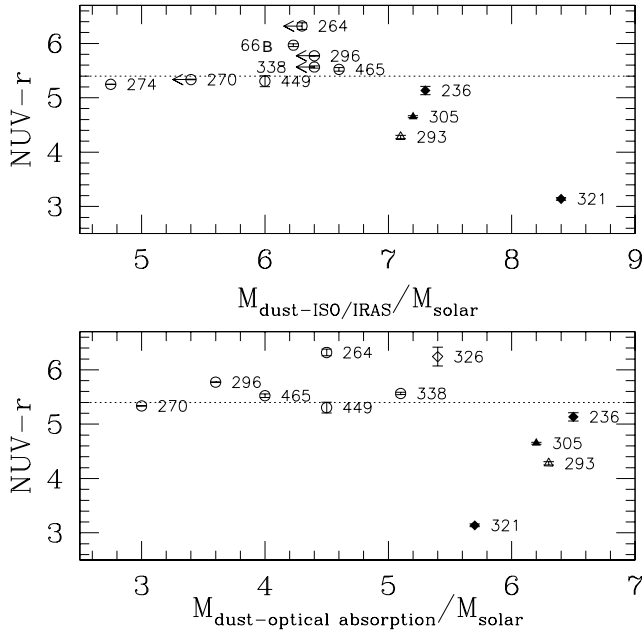,width=1\linewidth}}
\caption{The NUV-r color vs. dust mass (\citealt{dekoff00} and
  \citealt{muller04}) of the 3CR sample. Note that the HEG galaxies have dust
  content greater than the other objects. Symbols as in Fig. \ref{Lradio}.}
\label{dust}
\end{figure}

We now consider in more detail the properties and differences between the HEG
and LEG classes.  Both the UV emission and dust structures in HEG suggest that
these galaxies underwent a recent major merger and the highly chaotic and
unsettled morphologies imply an external origin for the dust
\citep{tremblay07}. The amount of dust derived by \citet{dekoff00}, using
optical absorption maps, and \citet{muller04}, from ISO and IRAS observations,
in HEG galaxies is also quantitatively larger than in LEG (see
Fig. \ref{dust}). This indicates that a significant amount of gas is provided
by the cannibalized object and appears to correspond to a ``wet'' merger.

Conversely, with one exception, we found no evidence of recent star formation
in the LEG galaxies.  Dust is observed in many of these galaxies, although
only in the form of circum-nuclear disks. While the formation mechanisms of
these structures are still largely unknown, \citet{temi07} proposed that the
accumulation of cold dust in the central regions of massive ellipticals is due
to mass loss from red giant stars in the galaxy core and is therefore of
internal origin. Their optical and UV morphologies are also relaxed and lack
clear signs of a recent merger. \citet{colina95} found evidence of subtle
isophotal disturbances in FR I galaxies, which could represent ``dry'' mergers
at a late stage.

\begin{figure*}
\centerline{
\psfig{figure=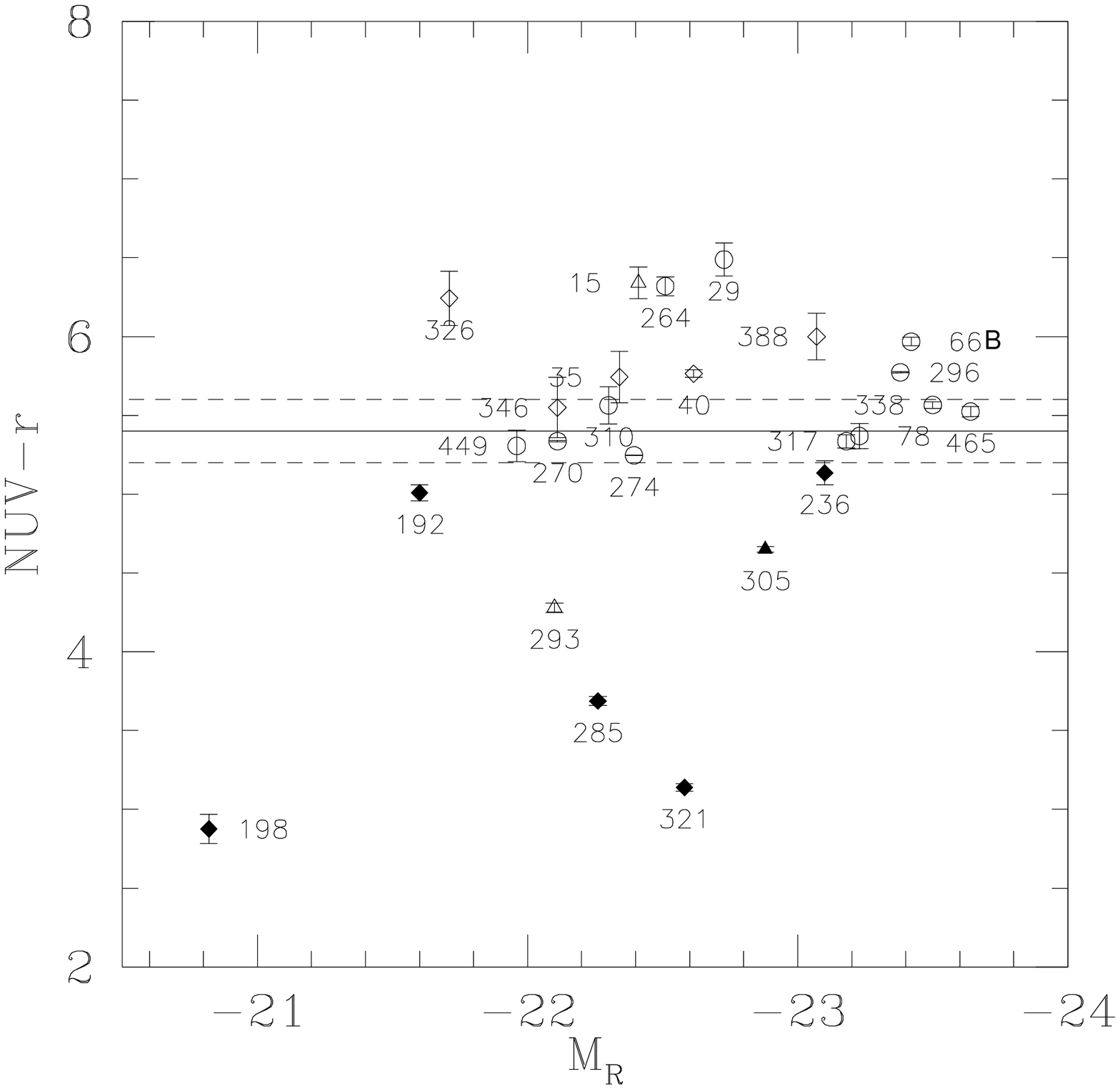,width=0.5\linewidth}
\psfig{figure=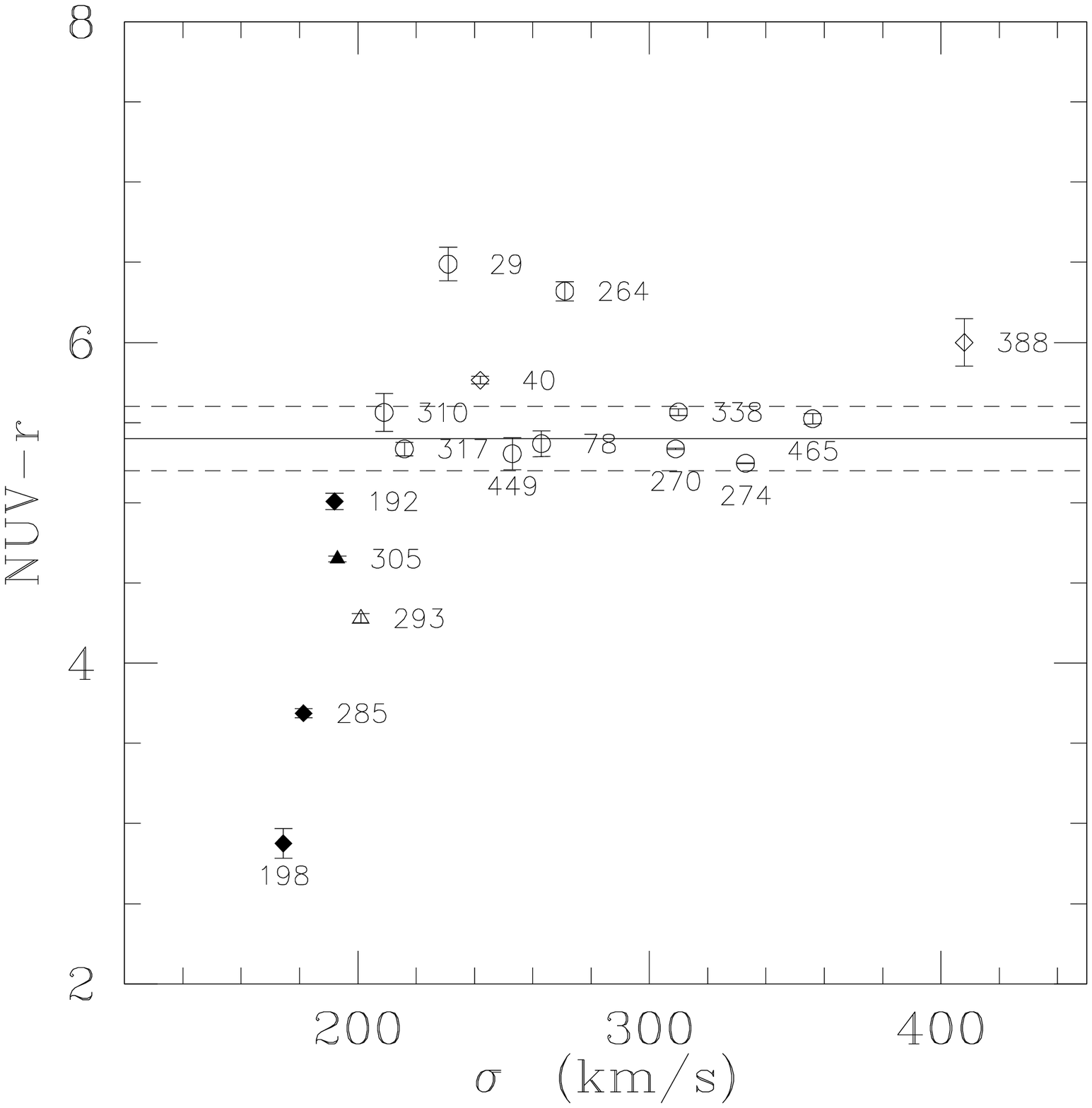,width=0.5\linewidth}}
\caption{Left panel: absolute magnitude M$_{R}$ vs NUV-r color. Right panel: stellar
  velocity dispersion vs NUV-r color. Symbols as in Fig. \ref{Lradio}.}
\label{mrc}
\end{figure*}

In terms of nuclear activity, HEG are associated to brighter AGN than
LEG. Within our sample, HEG have a median [O III] luminosity\footnote{ Narrow
emission-line luminosities, of in particular the [O III] line, provide robust
orientation-independent measures of the intrinsic AGN luminosity
\citep[e.g.][]{mulchaey94}.} higher by a factor of 30 with respect to LEG
\citep{chiaberge02}. This is reminiscent of the luminosity segregation between
high and low excitation AGN found by \citet{kewley06}, since they find that
the median [O III] luminosity for LINERs and Seyfert galaxies (the radio-quiet
analogous to LEG and HEG) differ by a factor of 16, Seyfert galaxies being
brighter.  We note that \citet{kewley06} also found that LINERs are older,
more massive, less dusty, and have higher velocity dispersions than Seyfert
galaxies, which is analogous to differences found between LEG and HEG.

In low luminosity radio galaxies or LEGs from a spectroscopic point of view,
the results by \citet{allen06,balmaverde08} indicate that the accretion rate
from the hot interstellar medium is proportional to the kinetic
output of their jets. This is an indication that this mode of accretion
(self-sustained by the interstellar medium) provides a sufficient energy input
to power their jets.  The higher AGN luminosity of HEG suggests that for these
objects there is the need of gas of external origin to satisfy their energy
requirements.

Another element of difference between HEG and LEG is their environment. It is
well known that FR~I are located in regions of higher galaxy density, often at
the center of galaxy cluster, than FR~II \citep{zirbel97}. Furthermore, among
the FR~II radio-galaxies, LEG are often found in clusters of galaxies
\citep{chiaberge00}, while HEG usually inhabit groups. The different galactic
environments can lead, at least from a statistical point of view, to the
different star formation manifestations between HEG and LEG.  In fact, in
groups of galaxies, the low relative velocity between objects and the higher
fraction of gas-rich galaxies increase the probability of major ``wet'' merger
events, whereas in cluster of galaxies, mostly ``dry'' mergers occur.  This
favors the conditions to trigger, via a ``wet'' merger, a powerful AGN and
significant star formation in the group galaxies and explains the different
environments between HEG and LEG.

The picture that emerges, considering i) NUV-r color, ii) UV and optical
morphology, iii) dust content, iv) nuclear luminosity, and v) environment is
that a recent major ``wet'' merger is needed to trigger the relatively
powerful AGN and copious star formation associated with the HEG.  Conversely,
in LEG galaxies the fraction of actively star-forming objects is not enhanced
with respect to that observed in quiescent galaxies. The accretion of the
interstellar medium also provides a sufficient energy input to power LEGs,
without the need for an external gas supply.

These results add to the list of differences between high and low excitation
radio-galaxies, which include the presence, only in HEG, of a prominent Broad
Line Region, an obscuring torus, and a radiatively efficient accretion
disk. All of these features are linked to their availability of sufficiently
high amounts of gas to form these structures.  It is also possible that HEG
and LEG are linked by an evolutionary sequence. A FR~I (or LEG) galaxy might
become a HEG when affected by a merger; at a later stage, when the fresh gas
is exhausted in the galaxy by star formation and accretion onto the SMBH, the
galaxy may revert to a quiescent LEG state.

We now compare our results with previous studies of the recent star formation
in radio-galaxies based the optical spectroscopy. Unfortunately, it is
difficult to perform a robust comparison of the fraction of actively
star-forming galaxies and its relation to the AGN properties. This is, in
part, due to the different selection criteria of the samples, the subjective
definition of the various sub-groups of radio galaxies, and the lack of
classification into the HEG and LEG classes for many objects studied in the
literature.  

Similarly, an object-by-object comparison of the level of AGN and nebular
continuum is not conclusive.  There is a strong band mismatch between our UV
data and the B band, where the contamination estimates are usually
made. Furthermore, the ground-based spectroscopy provides information that can
be matched reasonably only for our central aperture, and not for all
galaxies. In many cases we also excluded the central regions, where AGN
contamination is probably stronger, which are instead included in the
ground-based data.  Nonetheless, the levels of nebular continuum contamination
from spectroscopic analysis are comparable to our photometric estimates and is
tipically less 10\% (see Table \ref{contam}). Finally, we note that, quite
reassuringly, the two independent approaches converge to the same result for
all blue 7 objects in common.

We conclude the discussion with a further comparison with previous studies
of the NUV-r colors derived from the analysis of GALEX and SDSS images of
early-type galaxies. 

\citet{schawinski06} found that the fraction of RSF galaxies decreases with
increasing stellar velocity dispersion. They explained this trend to be an
AGN-feedback effect powered by supermassive black holes that halts star
formation in the most massive galaxies. A similar effect is observed for our
3CR sample (see Fig. \ref{mrc}, right panel), which appear to show a
connection between the stellar velocity dispersion and NUV-r color since the
bluest objects have preferentially low stellar velocity dispersion ($\sigma$
$\lesssim$ 250 km/s), while the red galaxies cover the entire range of
$\sigma$. However, a closer inspection shows that 4 of the galaxies with the
highest values of $\sigma$ are cD galaxies located at the center of clusters,
while one (namely 3C~270) is the dominant member of a rich group. Apparently,
the zone of avoidance (there are no blue galaxies at large $\sigma$) is mostly
driven by an environmental effect rather than a different level of AGN
feedback, and a blue NUV-r color is a manifestation of a recent merger.  A
quantitative comparison, however, is not straightforward due to the relatively
small size of our sample and the cross-calibration uncertainties, and would
require the selection of quiescent galaxies only in a rich environment.

The study of \citet{kauffmann06} was instead based on a volume-limited sample
of massive bulge-dominated galaxies, which also included AGN. Their GALEX-SDSS
color profiles demonstrated that the UV excess light is almost always
associated with an extended disk for galaxies with young bulges and strongly
accreting black holes. They suggested that the presence of an extended gas
structure is a necessary condition for AGN activity.  Unfortunately, the small
field-of-view of the STIS images used in our study does not allow us to
investigate directly whether low surface brightness UV disks are associated
with our radio-galaxies. Furthermore, the \citet{kauffmann06} sample cannot
be used as a reliable reference for our 3CR sources, since it is dominated by
galaxies of lower $\sigma$ and contains mostly radio-quiet AGN.

\section{SUMMARY AND CONCLUSIONS}
\label{summary}

We have analyzed images of 3CR radio galaxies for which both optical and UV
HST images are available. The sample is composed of 31 radio-galaxies, all but
one with z $<$ 0.1, representing $\sim$ 60 \% of all 3CR sources below this
redshift limit. We have excluded six objects, four are very highly nucleated
and two with data of too low signal-to-noise ratio, and have 25 remaining
objects. To perform a rigorous comparison with previous studies based on GALEX
(NUV band) and SDSS (r band) observations, we derived a cross-calibration
between GALEX, SDSS, and HST colors. We conservatively estimated that the
error associated with this procedure was $\lesssim$ 0.2 mag.

On the basis of the integrated colors and UV morphology, the galaxies of the
sample can be divided into 3 main categories: (1) the quiescent red galaxies
(14 objects) have ``red'' colors, with only diffuse UV emission, tracing the
optical light; (2) the ``blue'' UV-clumpy galaxies (7 objects) show a clumpy
UV morphology (extended over 5-20 kpc); (3) the UV-disky galaxies (4 objects)
have UV emission that is cospatial with their circum-nuclear (on a scale of
0.5-1 kpc) dusty disks.

To recognize the presence of recent star formation, it is necessary to account
for other sources of UV emission. We estimated the level of UV contamination
from emission lines and nebular continuum. In UV-disky galaxies, we found that
UV emission lines contributed $\sim$ 50 - 80 \% of the total UV flux within
the region coincident with the dusty disks, causing the observed blue NUV-r
color. For the 7 UV-clumpy galaxies, the emission line or nebular continuum
contamination were both negligible. There was a spatial association between UV
light and radio-axis only in one galaxy, which is a possible sign of
contamination from scattered nuclear light.  Summarizing, we confirm the
presence of a young stellar population in at least 6 of the 7 UV-clumpy
galaxies, in full agreement (on an object-by-object comparison) with results
obtained for ground-based optical spectroscopy.

The NUV-r color is not simply related to the radio power since, while the
bluest galaxies have all relatively large values of P$_{178 \rm MHz}$, at the
high end of radio luminosities we found many of the reddest galaxies.
Concerning the FR type, we found no blue galaxy among the FR~I sources, but
among the FR~II (or transiction FRI/II) galaxies there was an almost equal
number of ``blue'' and ``red'' galaxies.  The clearer association between
galaxy color and AGN type is related to the optical spectroscopic
classification into HEG and LEG: all six HEG are blue, while there is only 1
blue LEG out of 19 including all FR~I into this class.  The fraction of star
forming HEG is far in excess of that found by \citet{schawinski06} for their
sample of quiescent galaxies, while in LEG there appear to be even less
star-forming galaxies than in quiescent galaxies. These results can be
summarized as follows:
\begin{enumerate}
\item   {While all HEG have a blue UV-r color, the opposite is true for LEG,
with only one exception;}

\item {UV, optical, and dust morphology of HEG are highly chaotic and
  correspond to unsettled morphologies. In LEG, the UV emission is typically
  diffuse and traces the optical light (leaving aside the UV-disky galaxies
  contaminated by line emission), while dust (when present) is mostly arranged
  in small circum-nuclear disk structures;}

\item   {HEG have a higher dust content;}

\item   {HEG have higher nuclear luminosity.}
\end{enumerate}

All of these findings can be explained if for HEG we are seeing the effects of
a recent, major, ``wet'' merger. The fresh input of gas and dust causes both
the higher star formation rate and stronger nuclear activity.

The most favorable situation for the occurrence of a ``wet'' merger is in
groups of galaxies. This can explain the different environment between HEG,
usually located in groups, and LEG, more often found in clusters of galaxies.

Conversely, in LEG we did not find evidence for recent star formation (with
only one exception). A quantitative comparison with the star-formation fraction
of quiescent galaxies is not straightforward due to the relatively small size
of our sample and the cross-calibration uncertainties. However, we note that
the substantial tail of blue objects found in inactive early-type galaxies
is apparently not present for our objects, and this strongly disfavours the
possibility of enhanced star formation in LEG hosts.

It appears that there is no clear connection between the recent star
formation and the presence of low excitation AGN.  This is in agreement with the
 idea that in LEG the hot interstellar medium is able to substain the AGN
activity via quasi-spherical accretion. This process does not require an
external cold gas supply that might become detectable in a star formation event.

The results presented here demonstrate the potential of these studies for
investigating the triggering mechanism of nuclear activity and star formation
in radio galaxies, a method that can be adopted also for other classes of AGN.
However, this approach can not provide a quantitative estimate of the star
formation history, an essential ingredient in the study of the coupling
between the growth of galaxies and SMBH. This crucial information can only
be derived by complementing the existing UV images with further data, e.g. optical and UV
spectra, that can be used to constraints the age of the recently formed
stellar population.

\acknowledgements

We would like to thank the anonymous referee, Clive Tadhunter, 
and Marco Chiaberge for their very
useful comments and suggestions. The authors acknowledge partial
financial support by PRIN - INAF 2006 grant.

\bibliography{my}

\end{document}